\documentclass[fleqn,usenatbib]{mnras}

\usepackage{anyfontsize}
\usepackage[dvipsnames]{xcolor}

\usepackage[T1]{fontenc}

\DeclareRobustCommand{\VAN}[3]{#2}
\let\VANthebibliography\thebibliography
\def\thebibliography{\DeclareRobustCommand{\VAN}[3]{##3}\VANthebibliography}

\usepackage{graphicx}	
\usepackage{amsmath}	
\usepackage{natbib}
\usepackage{tablefootnote}

\newcommand{\teff}{$T_{\mathrm{eff}}$}

\newcommand{\msol}{M$_\odot$}
\newcommand{\rsol}{R$_\odot$}

\newcommand{\kalpha}{$k_{\alpha_{\mathrm{MLT}}}$}
\newcommand{\amlt}{$\alpha_{\mathrm{MLT}}$}
\def\orcid#1{\kern .08em\href{https://orcid.org/#1}{\includegraphics[keepaspectratio,width=0.7em]{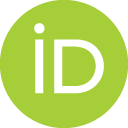}}}

\title[Isochrone fitting to M67]{Isochrone fitting to the open cluster M67 in the  era of Gaia and improved model physics}
\author[C. Reyes et al.]{
Claudia Reyes$^{1}$\orcid{0000-0001-9632-2706}, 
Dennis Stello$^{2,1}$\orcid{0000-0002-4879-3519}, 
Marc Hon$^{3,4}$\orcid{0000-0003-2400-6960}, 
Regner Trampedach$^{5}$\orcid{0000-0003-0866-6141}, 
Eric Sandquist$^{6}$\orcid{0000-0003-4070-4881}, and \newauthor Marc H. Pinsonneault$^{7}$\orcid{0000-0002-7549-7766}.\\
$^{1}$School of Physics, University of New South Wales, NSW 2052, Australia\\
$^{2}$Sydney Institute for Astronomy (SIfA), School of Physics, University of Sydney, NSW 2006, Australia\\
$^{3}$ Department of Physics and Kavli Institute for Astrophysics and Space Research, Massachusetts Institute of Technology, 77 Massachusetts Ave, Cambridge, \\ MA 02139, USA.\\
$^{4}$Institute for Astronomy, University of Hawaii, 2680 Woodlawn Drive, Honolulu, HI 96822, USA\\
$^{5}$Space Science Institute, 4765 Walnut Street, Boulder, CO 80301, USA\\
$^{6}$Department of Astronomy, San Diego State University, San Diego, CA 92182, USA\\
$^{7}$Department of Astronomy, The Ohio State University, Columbus, OH 43210, USA\\
}
\usepackage{newtxtext,newtxmath}
\date{Accepted XXX. Received YYY; in original form ZZZ}

\pubyear{2024}

\begin{document}
\label{firstpage}
\pagerange{\pageref{firstpage}--\pageref{lastpage}}
\maketitle

\begin{abstract}
The Gaia mission has provided highly accurate observations that have significantly reduced the scatter in the colour-magnitude diagrams of open clusters. As a result of the improved isochrone sequence of the open cluster M67, we have created new stellar models that avoid commonly used simplifications in 1D stellar modelling, such as mass-independent core overshooting and a constant mixing length parameter. 
This has enabled us to deliver a precise isochrone specifically designed for M67, available for download.
We follow a commonly-used qualitative approach to adjust the input physics to match the well-defined colour-magnitude sequence, and 
we test the model-predicted masses against a known eclipsing binary system at the main sequence turnoff of the cluster. Despite using improvements in photometry and stellar physics we cannot match the masses of both binary components with the same theoretical isochrone. A $\chi^{2}$-based isochrone fitting approach using our preferred input physics results in a cluster age of $3.95^{+ 0.16}_{- 0.15}$ Gyrs.

\end{abstract}




\section{Introduction}

The old open cluster M67 (NGC 2682) is known to be an important testbed for stellar evolution studies because of its low reddening, near-solar metallicity, and rich stellar population. Its turnoff stars fall         right in the mass range where we can infer how convective core size depend on stellar mass, making M67 particularly challenging, but also one of the only sites where we can use the colour-magnitude diagram to map out the convective core effects so readily. The colour-magnitude diagram of M67, together with stellar models, has therefore been used to determine the extent of convective core overshooting, to calibrate colour-temperature relations, and to test solar abundances. Examples include \citet{2004PASP..116..997V} and \citet{2007ApJ...666L.105V} who used $BV$ photometry from \citet{1993AJ....106..181M} and later \citet{2010ApJ...718.1378M} who used the higher precision $BV$ photometry from \citet{2004MNRAS.347..101S}. 

An unusually large fraction of binaries 
 and blue stragglers have been revealed through the study of M67's colour-magnitude diagram \citep{1971PASP...83..768S, 1986AJ.....92.1364M, 1991A&A...245..467M, 1993AJ....106.2441G, 2003AJ....125..246M}, and through spectroscopy \citep{2015AJ....150...97G, 2019ApJ...881...47L, 2021AJ....161..190G}. Various spectroscopic studies have also focused on the chemical abundances of M67 and have reported the signature of diffusion processes in the cluster's stars \citep{2018ApJ...857...14S, 2018MNRAS.478..425B, 2019ApJ...874...97S, 2019A&A...627A.117L, 2024MNRAS.529.2483B}.
In addition, the cluster's giants have been the targets of asteroseismic studies with both ground-based \citep{1993AJ....106.2441G, 2006MNRAS.373.1141S} and space-based telescopes \citep{2016ApJ...832..133S, 2024MNRAS.530.2810L}.

Since the first release of data from the Gaia mission \citep{2016A&A...595A...1G}, the observed colour-magnitude diagrams of clusters like M67 became significantly less scattered, and as a result new efforts to match theoretical isochrones to clusters have been made, including to M67 (\citet{2018A&A...616A..10G, 2018ApJ...863...65C, 2022A&A...665A.126N}).
However, none of the existing Gaia-era isochrones were explicitly created for M67 as the end goal, and as a result they do not match the Gaia data particularly well (Appendix \ref{sec:other_isos}, 
Figure~\ref{fig:other_isos}). Furthermore, they use the common approach of adopting a fixed mixing length parameter, \amlt, usually calibrated to the Sun, although there is no strong evidence suggesting \amlt\ is the same for stars of different evolutionary stages \citep{2020A&A...635A.176S}. This simplification can be expected to influence the temperatures and luminosities of stars with convective envelopes \citep{2023Galax..11...75J}.

In this paper, we aim to develop a new theoretical isochrone that incorporates the latest research on the physics governing stellar evolution: notably, model improvements based on hydrodynamical 3D simulations of convection \citep{2014MNRAS.442..805T, 2014MNRAS.445.4366T} while improving the fit to the colour-magnitude diagram. Such isochrone, 
validated by Gaia's M67 data, will provide an isochrone reference for future studies.
Additionally, we address the issue of the eclipsing binary WOCS11028, whose stellar components have highly accurate determinations of mass and radius and are found to likely not have interacted in the past  \citep[See][for an extensive discussion]{2021AJ....161...59S}. The problem is that no existing isochrone models simultaneously match these properties for both components.

Our approach is as follows: we begin using a traditional qualitative isochrone fitting approach \citep{2004PASP..116..997V, 2004ApJ...600..946P, 2018PASP..130c4204B} for empirical calibration of the input physics. Lastly, 
inspired by \citet{2023arXiv231116991L}, who explored a two-parameter grid to estimate age and extinction likelihoods in NGC 2516, we estimate age and distance modulus likelihoods for our models given our preferred input physics using a $\chi^{2}$ approach.

\section{The M67 Colour-Magnitude Diagram}
\label{sec:m67}

\subsection{Membership sample}
\label{sec:sample}
\begin{figure*}
\includegraphics[width=\textwidth]{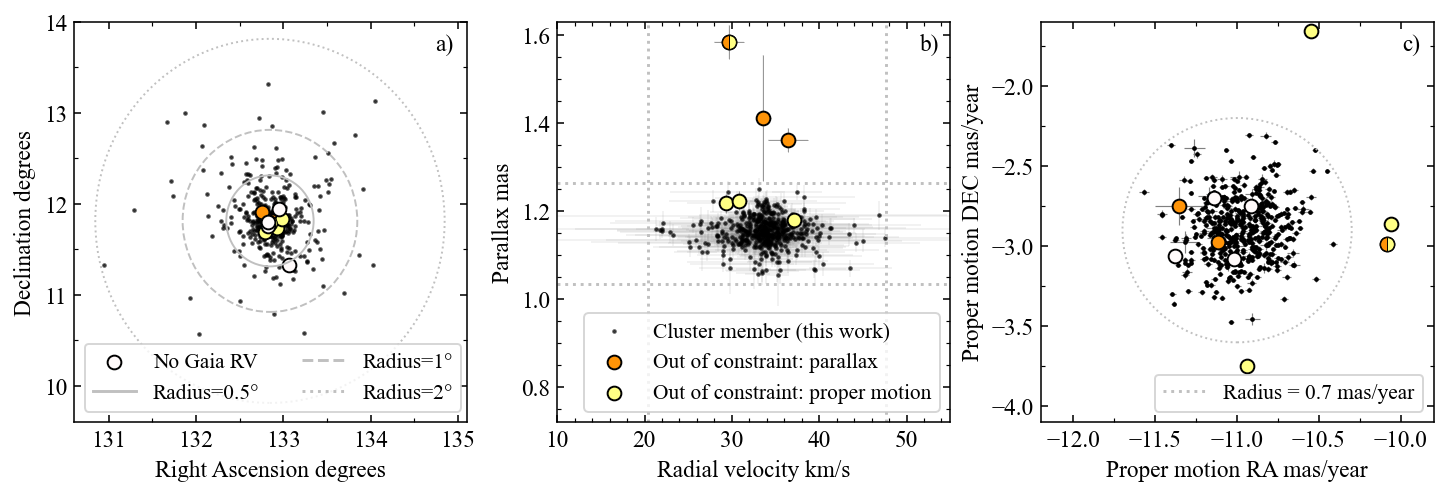}
\caption{In small black symbols are cluster members as per our selection using DR3 values. Large circles are the ten stars we added because they were deemed cluster members by \citet{2015AJ....150...97G} despite not being in our Gaia-based selection. Of those: (1) white circles show stars with no Gaia DR3 radial velocities available, (2) orange circles are stars that missed our cut in parallax, 
and (3) yellow circles are those outside our proper motion cut. 
The star shown in half-orange and half-yellow, WOCS 7004, falls outside of our proper motion and parallax cuts.}
\label{fig:Membership_p1.png}
\end{figure*}

\begin{figure*}
\includegraphics[width=\textwidth]{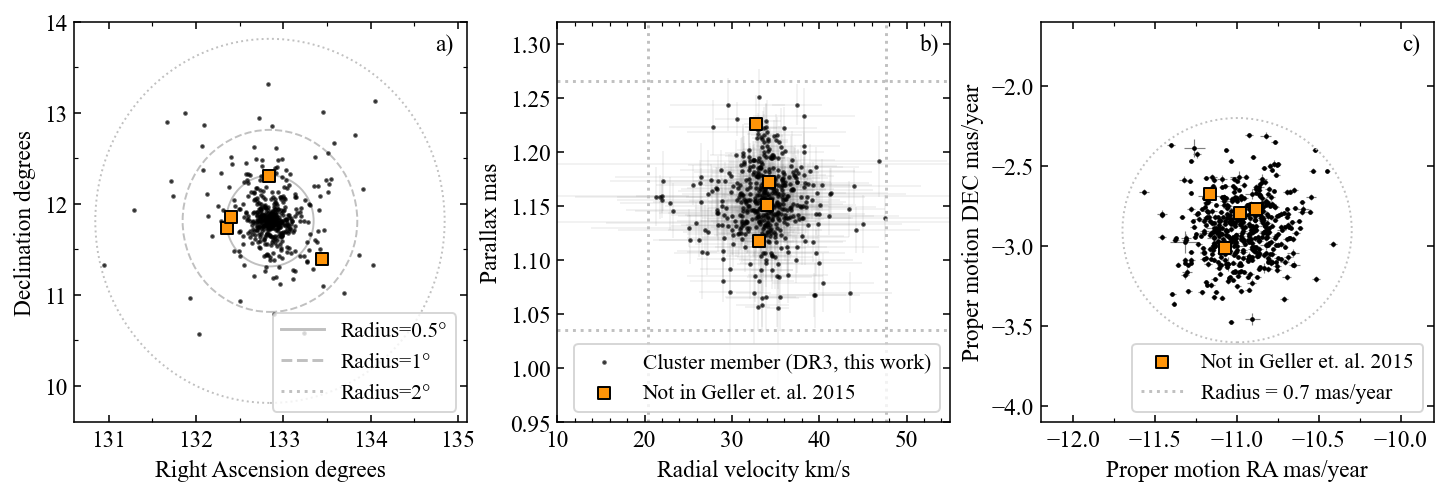}
\caption{The three panels show stars considered cluster members according to our Gaia-based selection. Stars in black symbols are also part of the membership sample by \citet{2015AJ....150...97G}, while orange squares show the four red giant stars that were not considered cluster members by \citet{2015AJ....150...97G}.}
\label{fig:membership2}
\end{figure*}

We determined cluster membership using Gaia DR3 \citep{2022yCat.1355....0G} from stars within 2 degrees of the centre of the cluster at $\mathrm{RA}=132.85^{\circ}$, $\mathrm{Dec}=11.81^{\circ}$ (Figure~\ref{fig:Membership_p1.png}a), selecting the overdensity of stars with radial velocities within $34 \pm 13.6\,\mathrm{km/s}$
and parallax $1.15\, \pm 0.12\, \mathrm{mas}$ (Figure~\ref{fig:Membership_p1.png}b), and proper motions within $0.7\, \mathrm{mas/year}$ from $\mu_{\alpha}=-11.0\, \mathrm{mas/year}$ and $\mu_{\delta}=-2.9\, \mathrm{mas/year}$ in the proper-motion space (Figure~\ref{fig:Membership_p1.png}c). To boost the number of stars along the less populated giant branch, we further added ten giant stars with secure kinematic membership from ground-based radial velocities by \citet{2015AJ....150...97G} that: (1) did not have DR3 radial velocities available (white circles in Figures \ref{fig:Membership_p1.png}a and \ref{fig:Membership_p1.png}c), or (2) narrowly missed our Gaia-based cuts in parallax or proper motion (orange and/or yellow circles in Figure~\ref{fig:Membership_p1.png}), but were later confirmed by colour-magnitude diagram position to belong to the cluster's standard-evolution single-star sequence. In total, our selection resulted in 488 cluster members: 22 non-standard evolution stars such as blue stragglers, yellow giants, and sub-subgiants; 361 main sequence stars, 63 subgiants, 35 red giant branch stars, and 7 red clump stars.
This number includes four new stars that were not identified by \citet{2015AJ....150...97G} as M67 members, but met all of our selection criteria for cluster membership, as shown in Figure~\ref{fig:membership2} with yellow squared symbols. Table~\ref{tab:table_membership} lists the stars mentioned above, along with their corresponding IDs from various catalogues.

\begin{table*}
    \centering
    \begin{tabular}{ccccccc}
Sanders\footnote{Catalogue } & WOCS\footnote{Catalogue \citet{2008A&A...484..609Y}} & EPIC\footnote{\citet{2017yCat.4034....0H}} & Gaia DR3\footnote{\citet{2022yCat.1355....0G}} & In DR3 selection & Geller'15 member & Comment \\
\hline
S 1016 & 2004 & 211410817 & 604917594995372544 & No & Yes & No RV available \\
S 978 & 1008 & 211407537 & 604911375882674560 & No & Yes & No RV available \\
S 1319 & 5021 & 211420451 & 604923470510324224 & No & Yes & No RV available \\
S 1557 & 1065 & 211380313 & 604696730596495744 & No & Yes & No RV available \\
S 1264 & 3004 & 211411629 & 604917835513458688 & No & Yes & Missed cuts in Parallax \\
S 806 & 5016 & 211417812 & 604969061587592064 & No & Yes & Missed cuts in Parallax \\
S 1221 & 2014 & 211406541 & 604904503934969856 & No & Yes & Missed cuts in Proper Motion \\
S 961 & 2017 & 211403555 & 604910001493155584 & No & Yes & Missed cuts in Proper Motion \\
S 1463 & 4017 & 211413064 & 604918522708195584 & No & Yes & Missed cuts in Proper Motion \\
S 1000 & 7004 & 211409644 & 604917320117398528 & No & Yes & Missed cuts in PLX + PM \\
S 258 & 1054 & 211414351 & 604965664269158656 & Yes & No & Red Giant, single star  \\
S 1135 & 2059 & 211443624 & 605015309794935552 & Yes & No & Red Giant, single star \\
S 247 & 5059 & 211406144 & 598959032246073216 & Yes & No & Red Giant, binary star \\
S 2000 & - & 211384259 & 604689308892956416 & Yes & No & Red Giant, No WOCS ID \\
\hline	
\multicolumn{7}{l}{$^{1}$\citet{1977A&AS...27...89S}, $^{2}$\citet{2008A&A...484..609Y}, $^{3}$\citet{2017yCat.4034....0H}, $^{4}$\citet{2022yCat.1355....0G}}\\

    \end{tabular}
    \caption{The 10 stars from Figure~\ref{fig:Membership_p1.png} and the 4 stars from Figure~\ref{fig:membership2}, along with their commonly used names.}
    \label{tab:table_membership}
\end{table*}

\subsection{Observed Colour-Magnitude Diagram}
\label{sec:cmd}

\subsubsection{Photometry}
\label{sec:photometry} 
We compared M67 photometry from Gaia DR2 \citep{2018A&A...616A...1G} and DR3 \citep{2022yCat.1355....0G} and found a similar scatter in the colour-magnitude sequence of the two systems. Therefore, either system could be used for isochrone fitting without precision loss. We decided to use Gaia DR2 because the deconvolution photometry of the eclipsing binary WOCS 11028 \citep{2021AJ....161...59S} at the cluster's turnoff is not directly available in Gaia DR3 magnitudes, and we want to use the binary system during the isochrone fitting (section \ref{sec:isochrone}).

\subsubsection{Isochrone fitting sample}
\label{sec:fitting_sample} 
We exclude from our isochrone-fitting sample those stars identified as binary, likely binary, stragglers, or possibly contaminated by a nearby source according to \citet{2015AJ....150...97G}, and stars from the Gaia DR3 non-single star catalogue \citet{2022arXiv220605595G}. Following \citet{2019A&A...628A..35K}, we also remove stars with a renormalised unit weight error (RUWE) that exceeds 1.2 in DR2, all of which are likely to fall far from the tight single-star cluster sequence in the colour-magnitude diagram. This selection results in a sample of 369 stars (small grey and black dots in Figure~\ref{fig:reddening}). The confidence level that this sample is free of multiple star systems mainly depends on the accuracy of the radial velocities from \citet{2015AJ....150...97G}, which degrade toward stars fainter than $12^{\mathrm{th}}$ magnitude in $V$ (see their Figure~4). To take care of the remaining stars that clearly belong to the binary main sequence by their position in the colour-magnitude diagram, we made a further manual cut that brings the isochrone fitting sample to the 309 stars shown in black in Figure~\ref{fig:reddening}.

\subsubsection{Reddening}
\label{sec:reddening} 
\begin{figure}
\centering
\includegraphics[width=\columnwidth]{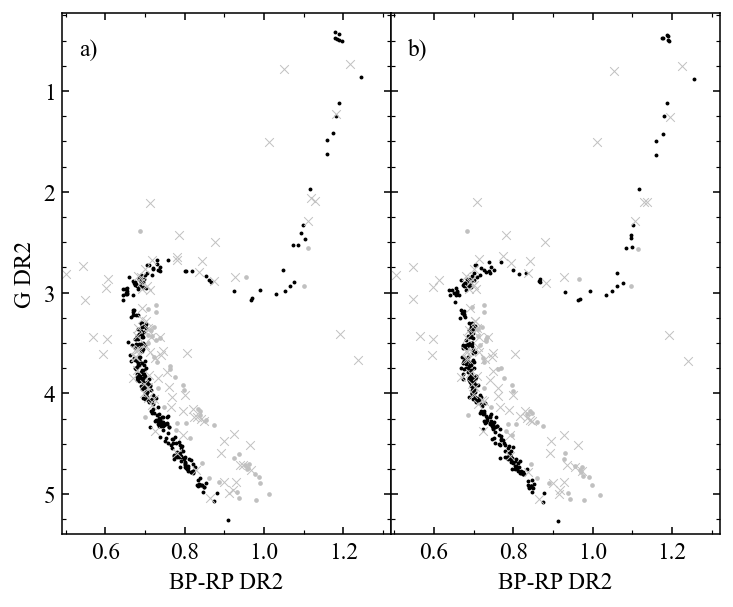}
\caption{ Black symbols represent stars in the isochrone fitting sample, grey `x' symbols show presumed binaries or presumed products on non-standard evolution, and grey points show stars we later removed from the isochrone-fitting sample (see Section \ref{sec:fitting_sample})
(a) M67 colour-magnitude diagram (absolute magnitudes) after uniform reddening (2.72 x 0.041 G-magnitudes) and extinction (0.06 BP-RP magnitudes) corrections. (b) Applying differential corrections based on dustmaps and stellar coordinates.}.
\label{fig:reddening}
\end{figure}

For differential de-reddening of the photometry, we sought the latest version of the Bayestar 3D dust maps \citep{2019ApJ...887...93G}. However, we found that these maps lacked resolution and did not behave smoothly in the direction of M67 (See Appendix \ref{sec:map}). Thus, we performed de-reddening on a star-by-star basis using 2D dust maps by \citet{2016A&A...594A..13P}, and
 calculated individual extinction coefficients using the formulae from \citet{2018A&A...614A..19D} with polynomial coefficients as calculated by \citet{2018A&A...616A..10G} assuming $R_{V}=3.1$. 
The averages of the full membership sample of 488 stars are $E(BP-RP)=0.059\pm0.007$ and $A_{G}=0.104\pm0.012$ magnitudes. Assuming $E(B-V)=0.041$ as in \citet{2007AJ....133..370T}, our values are in agreement with \citet{2019AJ....158..138S} who found the relations $E(BP-RP)=1.31\cdot E(B-V)$ and $A_{G}=2.72\cdot E(B-V)$. Even in a low reddening cluster like M67, 
applying individual corrections for dereddening and extinction reduced the scatter of the colour-magnitude sequence. This was most noticeable around the hook and the lower red giant branch, but also along the main sequence (compare Figures~\ref{fig:reddening}a and b).

\subsubsection{Distance}
\label{sec:distance} 
We derived the distance modulus from our sample using Gaia DR3 parallaxes, which are shown to have increased precision by 30\% over DR2 parallaxes \citep{2021A&A...649A...1G}, and we applied zero point offsets calculated for each star following \citet{2021A&A...649A...4L}, who found that zero-points depend on stellar magnitude, colour, and ecliptic latitude.
The mean zero-point corrected parallax derived this way is $\bar{\omega} = 1.195 \pm 0.027$ mas corresponding to a distance modulus $=9.614\pm 0.049$ magnitudes. 
This is a smaller distance modulus than the values of 9.726 and 9.730, derived by \citet{2018A&A...616A..10G} and \citet{2018ApJ...863...65C}, respectively, and also smaller than 9.690 adopted by \citet{2022A&A...665A.126N}, 
but close to 9.630 by \citet{2021AJ....161...59S}.  These four studies base their distance modulus calculation on Gaia DR2 parallaxes, but the latter two apply a parallax zero-point offset. \citet{2022A&A...665A.126N} correct their value applying directly the offset of $-30\mu \mathrm{as}$ found by \citet{2018A&A...616A...2L}, while \citet{2021AJ....161...59S} estimate an offset of $-54\mu \mathrm{as}$ considering the results by \citet{2018ApJ...862...61S}, \citet{2019ApJ...878..136Z}, and \citet{2019MNRAS.487.3568S} in addition to \citet{2018A&A...616A...2L}.

\subsubsection{Metallicity}
\label{sec:feh}
\begin{figure}
\includegraphics[width=\columnwidth]{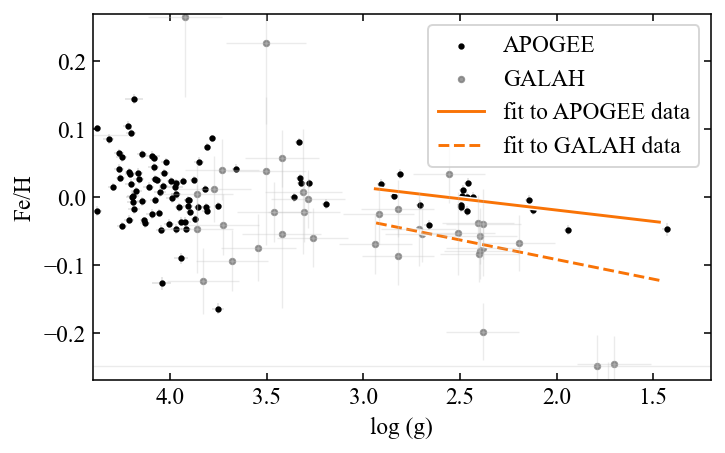}
\caption{Metallicities and $\log (g)$ by APOGEE DR17 and GALAH DR3 for stars overlapping with our 488 cluster members. The orange lines show fits to the metallicities of the stars with $\log(g)<3.0$ for both APOGEE (solid line) and GALAH (dashed line). In the latter, two extreme outliers at low $\log (g)$ and low [Fe/H] were excluded from the fit. All but one of the stars in the GALAH sample are also in the APOGEE sample}. 
\label{fig:apogee2}
\end{figure}

M67 has been observed extensively by spectroscopic surveys and it has even been used as a calibration cluster for APOGEE \citep{2022AJ....163..229S}. Naturally, we turned to this survey for updated M67 metallicity values. Figure~\ref{fig:apogee2} shows [Fe/H] from APOGEE DR17 for our cluster members (black dots).
We noticed a decreasing metallicity trend with decreasing $\log(g)$, affecting the sample's most evolved giants. This was also observed in other open clusters from DR17 (See Appendix \ref{sec:apogee}). For comparison, we looked at GALAH DR3 results of M67 (grey dots) \citep{2021MNRAS.506..150B}. Even though the GALAH data has higher dispersion and uncertainties, it also showed a descending trend with $\log(g)$, albeit more pronounced.
This common trend is likely systematics from not using a full 3D-NLTE approach in the spectroscopic analysis, affecting stars with $\log(g)\approx3$-2.5 or lower. No physical mechanism intrinsic to the stars can explain such metallicity decline in stars undergoing convective mixing in the atmosphere. This issue could lead to an error in [Fe/H] of up to 1 dex in some cases (for example, NGC 2632 in Figure~\ref{fig:apogee1}). Therefore, we caution against relying on giant star metallicities from non-3D-NLTE spectroscopic results, 
and we turn to recent M67 literature to find an appropriate [Fe/H] value for our models.

The average metallicity across all SDSS DR12 \citep{2015ApJS..219...12A} M67 observations was reported as $\mathrm{[Fe/H]}=0.08 \pm 0.03$ by \citet{2016ApJ...832..133S}, with solar reference from \citet{2005astro.ph.10377A}. Two recent studies have used high-resolution spectroscopic observations to examine the chemical abundance trends of stars along the cluster isochrone to investigate the effects of atomic diffusion.
\citet{2019A&A...627A.117L} used Keck/HIRES spectra to obtain high-precision atmospheric parameters and elemental abundances, and \citet{2019ApJ...874...97S} based their work on APOGEE DR14 data \citep{2018AJ....156..125H}. They both found that stellar models needed to have a [Fe/H] between 0.0 and 0.1 to match the detailed chemical abundance patterns, and both used MIST tracks with diffusion \citep{2016ApJ...823..102C}, and the solar mixture by \citet{2009ARA&A..47..481A} (the AGSS09 scale).
\citet{2019ApJ...874...97S} found no significant abundance difference in M67 stars of the same evolutionary class, but found abundance differences between stars in different evolutionary stages of up to $\sim 0.5 \mathrm{dex}$. They found that the observed abundance trends can be explained by diffusion processes, and estimated a reduction in the efficiency of gravitational settling of 15\%.

\section{Isochrone Models}
\label{sec:isochrone} 

We produce custom isochrones tailored to M67 using MESA version 23.05.1 \citep{2011ApJS..192....3P, 2013ApJS..208....4P, 2015ApJS..220...15P, 2018ApJS..234...34P, 2019ApJS..243...10P, 2023ApJS..265...15J}. 
To transform the theoretical isochrones into magnitudes that allow for direct comparisons with observations, we use bolometric correction tables from the MIST project\footnote{https://waps.cfa.harvard.edu/MIST/model\_grids.html}, which are based on the grid of stellar atmospheres and synthetic spectra described by \citet{2016ApJ...823..102C}. 
Our approach for selecting model parameters to match M67, resulting in our final preferred isochrone, named Isochrone A\footnote{Isochrone A and Inlists available on Zenodo \url{https://zenodo.org/records/12616441}}, is described in the following. 
Table~\ref{tab:isoa} summarises Isochrone A's parameters.
\begin{table}
    \centering
    \begin{tabular}{ll}
        \hline
        Parameter & Isochrone A\\
        \hline
        Age & 3.95 Gyrs\\
        Distance modulus & 9.614\\
        Nuclear reaction network & h1, he3, he4, c12, n14, o16, ne20, mg24\\
        & (MESA's basic.net)\\
        Solar mixture  & AGSS09 {\citep{2009ARA&A..47..481A}}\\
        Z & 0.016\\
        Y & 0.267\\
        $\Delta$Z/$\Delta$Y & 1.2\\
        EoS &   HELM \citep{2000ApJS..126..501T}\\
         & + Skye \citep{2021ApJ...913...72J}\\
          & +  FreeEOS \citep{FreeEOS}\\
          & + OPAL \citep{2002ApJ...576.1064R}\\
          & + SCVH \citep{1995ApJS...99..713S}\\
        Interior opacities & OPAL \citep{1993ApJ...412..752I, 1996ApJ...464..943I}\\
        Opacities & AESOPUS {\citep{2009A&A...508.1539M}}\\
        Atmospheres & T($\tau$), varying, Trampedach solar\\
         & \citep{2014MNRAS.442..805T, 2021RNAAS...5....7B}\\
        Overshooting & Exponential, increasing with mass. \\
        \amlt\ & Varying with \teff\ and $\log(g)$,\\
         & follows 3D grid \citep{2014MNRAS.445.4366T}\\ 
        Scaling factor \kalpha & 1.11\\
        Mass loss & No\\
        MLT\_option & Henyey\\
        Convection & Schwarzschild criterion\\
        Diffusion & Yes\\
        \hline
    \end{tabular}
    \caption{Summary of parameters in MESA models for the isochrone shown in Figure~\ref{fig:isochrone}. The isochrone is presented in \ref{tab:isocsv}}.
    \label{tab:isoa}
\end{table}

\subsection{Initial chemical composition}
\label{sec:chemical_composition} 

Guided by the results detailed in \ref{sec:feh} we decided to adopt the AGSS09 scale, $\mathrm{[Fe/H]}=0.05$, and $Z=0.016$, 
implying a helium enrichment law of $\Delta Y/\Delta Z= 1.2$ when primordial helium is taken as $Y_{\mathrm{p}}=0.248$. While helium is not solar-scaled, similar values of $\Delta Y/\Delta Z$ have been used to match models to data
in other open clusters: 1.4 was found for NGC 6791 \citep{2012A&A...543A.106B}, 1.2 for the Hyades \citep{2021A&A...645A..25B}, and an average of 1.25 for NGC 6811 \citep{2016ApJ...831...11S}. 
\citet{2019ApJ...874...97S} predicted [Fe/H] $\sim 0.05$ (our chosen primordial metallicity) for M67 solar analogues using their estimated diffusion factor, which is consistent with our models if we consider our lower than solar initial helium, and their solar scaled models. 

Our choice of $Z$ achieves good colour agreement with the main sequence for ages within the range quoted in literature: 3.5 - 4.3 Gyrs \citep{2004PASP..116..997V, 2009ApJ...698.1872S, 2016ApJ...823...16B}, while the chosen helium content, $Y=0.267$, results in slightly more massive stars for a given luminosity compared to solar-scaled models. This becomes relevant when taking into account the eclipsing binary WOCS 11028 (Section \ref{sec:binary}).

\subsection{Model atmospheres}
\label{sec:atmosphere} 
One-dimensional stellar evolution codes like MESA include simplifications in their representation of the outer layers of a star, such as  
using a single and fixed value of the convective mixing length \amlt, which is commonly anchored to the Sun. In our models, we seek to incorporate more realistic values of \amlt. This, we take from interpolations to the grid of hydrodynamical 3D simulations of stellar convection -- the 3D grid -- by \citet{2014MNRAS.445.4366T} evaluated at every step according to $\log(g)$ and \teff\ with a scaling factor \kalpha, following \citet{2018MNRAS.478.5650M}, except that we calibrate \kalpha\ to M67 rather than to the Sun (Section \ref{sec:mixing}). 

To ensure the correct photospheric transition to the outer part of the models, we follow the formulation of generalised Hop-functions for atmospheres by \citet{2014MNRAS.442..805T} from 3D simulations of convection in deep stellar atmospheres: ($T(\tau)$=$\texttt{Trampedach\_solar}$) as implemented by \citet{2021RNAAS...5....7B}, and we modify the radiative gradient so that regions of low optical depth have a temperature that follows this $T(\tau)$ relation.
The results of applying the improvements described above to a base form of Isochrone A can be observed in Appendix~\ref{sec:other_isos}, by comparing Figures~\ref{fig:iso_other}a and b.

While studying the effects of varying model parameters, we saw that the selection of low-temperature opacities had a significant impact on the temperatures and luminosities of the models. As a result, we have opted to use low-temperature opacities from the AESOPUS (Accurate
Equation of State and Opacity Utility Software) tables because the AESOPUS code \citep{2009A&A...508.1539M} accounts for the processing of CNO in giant stars while the MESA default tables FA05 \citep{2005ApJ...623..585F} do not.
The effect of additionally changing from FA05 to AESOPUS opacity tables can be observed in Appendix~\ref{sec:other_isos}, by comparing Figures~\ref{fig:iso_other}b and c.

\subsection{Overshooting}
\label{sec:overshoot_massloss}
We implement exponential diffusive overshoot \citep{2000A&A...360..952H} at the Hydrogen core  
with a strength set to increase gradually with mass, following the expression proposed by \citet{2006ApJS..162..375V}:
\begin{equation}
\label{eq:overshoot}
    f_{\mathrm{ov}} = f \left[ 1 - \mathrm{cos}\left(\pi \frac{M_{\star}-M_{\mathrm{full\_off}}} {M_{\mathrm{full\_on}}-M_{\mathrm{full\_off}}} \right) \right]
\end{equation}
\noindent where $M_{*}$ is the mass along the isochrone, $M_{\mathrm{full\_off}}=1.10\,\mathrm{M}_{\odot}$, $M_{\mathrm{full\_on}}=1.80\, \mathrm{M}_{\odot}$, which provides a smooth transition between 0 and $f_{\mathrm{ov}}$.
We set the parameter $f_{0}$ as $f_{\mathrm{ov}}/2$, where $f_{0}$ is a distance from the edge of the convective zone where the diffusion mixing coefficient is measured. This coefficient is then multiplied by $e^{-2z/f_{\mathrm{ov}}H\!p}$, where $H\!p$ is the pressure scale height,
so that the mixing efficiency declines exponentially into the radiative zone with
distance $z$ from the convective boundary.

While it is common to define overshoot in terms of $H\!p$, this poses a problem when $f_{\mathrm{ov}}$ is kept fixed for all masses because $H\!p\!\rightarrow\!\infty$ when the convective core radius $\rightarrow\!0$. By default, MESA limits $f_{\mathrm{ov}}H\!p$ to the size of the convective zone, however, this could still result in unphysically large convective cores in low mass stars, as, according to \citet{1992A&A...266..291R} overshooting should not expand the convective core more than 18\% of its `no-overshoot' size.
A mass-dependent overshoot like the one from Equation (1) helps overcome this problem. 
Another common criterion that avoids the issue is to express the overshoot extent as a fraction of the core radius.
The effect of the change from a fixed overshoot to our adopted mass-dependent values can be seen in Appendix~\ref{sec:other_isos}, by comparing the black and green curves in Figure~\ref{fig:iso_other}c.

\subsection{Mass Loss}
We do not have a complete understanding of the mass-loss processes operating during the red giant phase \citep{Karakas2017}. \citet{2016ApJ...823..102C} selected a Reimers parameter $\eta=0.1$ for use in their MIST models based on the initial-final mass relation in the Magellanic Clouds.
\citet{2012MNRAS.419.2077M} found from asteroseismology that the mass difference between red giant branch and red clump stars of the metal-rich open cluster NGC 6791 can be described with a Reimers parameter in the range $0.1<\eta< 0.3$, lower than predicted by the commonly used prescription by \citet{1975psae.book..229R}. They also found that the near-solar-metallicity open cluster NGC 6819 is compatible with no mass loss. 
\citet{2016ApJ...832..133S} measured the masses of M67 giants using asteroseismology and found no significant mass difference between the red clump and the red giant branch. 
Considering all these results, and especially the evidence from M67 asteroseismology, we decided to run models without mass loss. 
Regardless, mass loss is unlikely to significantly affect our M67 isochrone fit because nearly all mass loss occurs at the very end of the red giant branch, and red clump stars will be excluded from the final fit (see Section \ref{sec:chisq}).

\subsection{Empirical Calibrations}
Bringing together our model prescriptions as described above and the extinction-corrected M67 colour-magnitude diagram from Section \ref{sec:cmd}, we can now fine-tune the mixing length and overshooting. 
\begin{figure*}
\includegraphics[width=\textwidth]{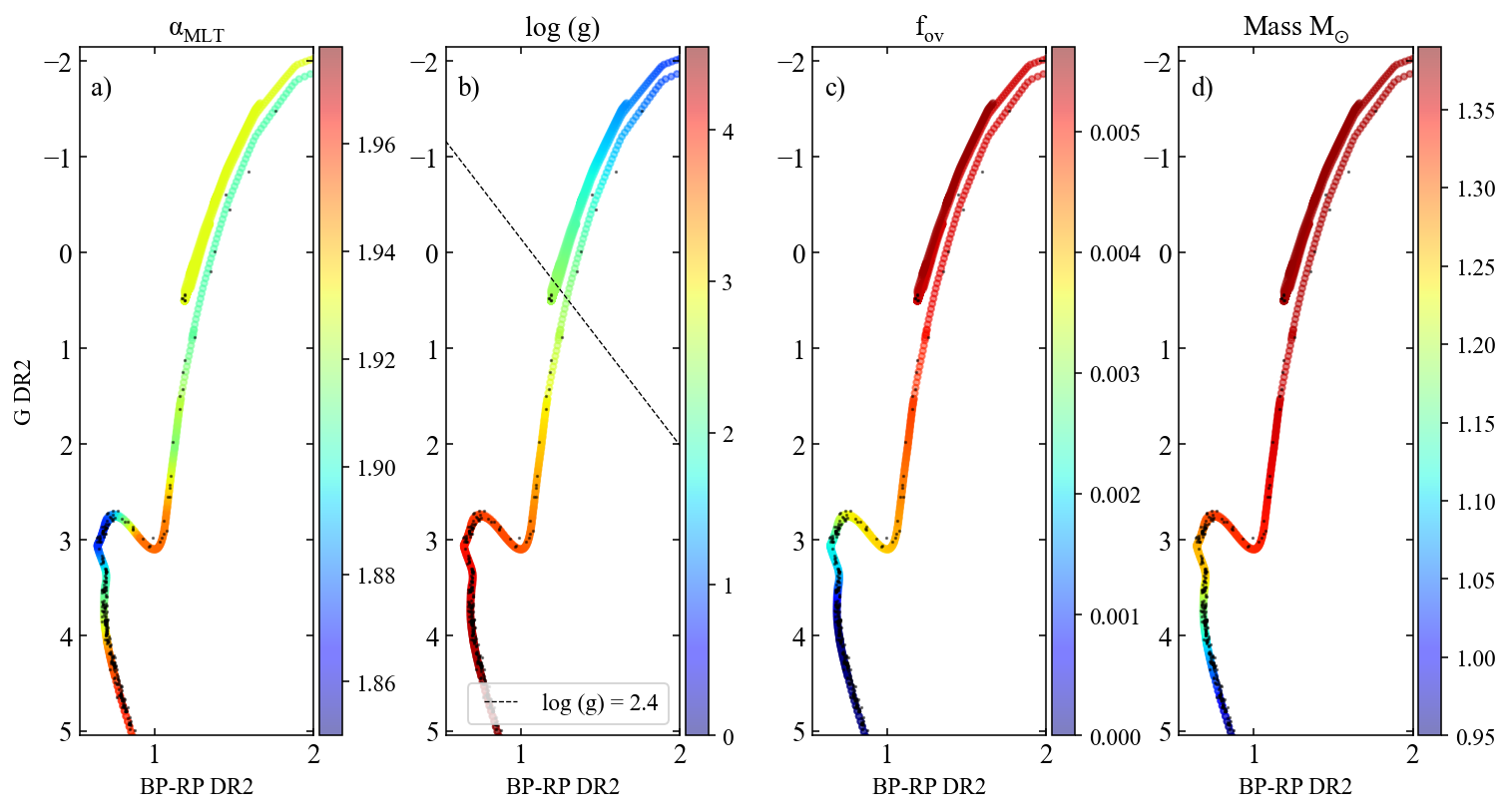}
\caption
{Isochrone A and M67 presumed single stars in absolute magnitude and colour-coded by (a) \amlt, (b) $\log(g)$, (c) overshooting parameter $f_{\mathrm{ov}}$, and (d) stellar mass. The dashed line in b) crosses the isochrone where the \amlt\ 3D grid stops at $\log(g)$ 2.4. After that point in the evolution up the RGB, we set our models to maintain the last measured \amlt\ value and after helium ignition \amlt\ is set to 1.93. }
\label{fig:isocolor}
\end{figure*}

\subsubsection{Mixing Length Parameter}
\label{sec:mixing}
Figure~\ref{fig:isocolor}a shows Isochrone A colour-coded by our adopted mixing length, resulting from applying the scaling factor \kalpha = 1.11 to the 3D grid. Because the 3D grid stops below $\log(g) \sim$2.4 --segmented line in Figure~\ref{fig:isocolor}b--, 
we keep the last interpolated grid value for the rest of the evolution up the giant branch, but doing so puts the isochrone's red clump at lower temperatures than observed from the cluster's clump stars. To resolve this issue, we decided to set \amlt\ to 1.93 after the helium flash, which makes Isochrone A agree with the observations of the red clump. However, this does not necessarily mean that 1.93 is the correct mixing length value for stars with \teff\ and $\log(g)$ similar to the M67 clump stars. It merely expresses our ignorance of the behaviour of \amlt\ after the 3D grid stops and also compensates for other shortcomings that our helium core burning models might have that would affect \teff.

\subsubsection{Overshooting}
\label{sec:overshoot}

\begin{figure}
\centering
\includegraphics[width=\linewidth]{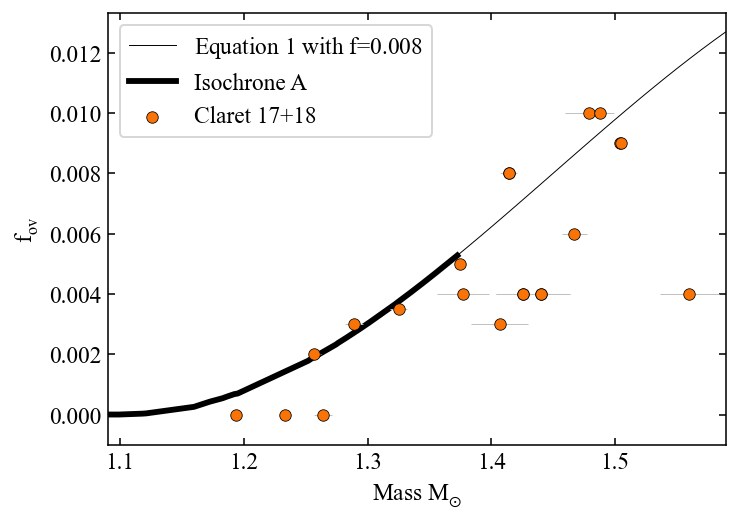}
\caption{Comparing our mass-dependent overshoot used in Isochrone A with independent modelling of eclipsing binaries \citep{2017ApJ...849...18C, 2018ApJ...859..100C}. The dark section of the curve indicates the mass range along Isochrone A.}
\label{fig:overshoot_other}
\end{figure}

The effect of the overshooting parameter is seen primarily near the isochrone's turnoff due to the turnoff's sensitivity to the size of the convective core and how the size of the convective core depends on stellar mass. 
Figures \ref{fig:isocolor}c and d show Isochrone A colour-coded by $f_{\mathrm{ov}}$ and stellar mass, respectively. 
Figure~\ref{fig:overshoot_other} shows our mass-dependent overshooting parameter $f_{\mathrm{ov}}$, compared to the reported $f_{\mathrm{ov}}$ by \citet{2017ApJ...849...18C, 2018ApJ...859..100C} in orange symbols, where models using exponential overshooting with diffusion are fitted to eclipsing binary masses. 

There have been several recent studies on overshooting, but they often use different methods that cannot be directly compared to our calibration. For example, \citet{2024arXiv240212461L} applies `step' overshoot instead of exponential, and maintains a fixed value of $f_{0}$ while they vary $f_{\mathrm{ov}}$ and take care of the issue of large cores at low masses mentioned in Section~\ref{sec:overshoot_massloss} by switching from units of $H\!p$ to units of core radius if $H\!p$ is larger than the core radius.
For an in-depth review of their results in the context of results from various overshooting studies, including \citet{2017ApJ...849...18C, 2018ApJ...859..100C}, we refer the reader to \citet{2024arXiv240212461L}.

\subsection{Eclipsing Binary WOCS 11028}
\label{sec:binary}
The discovery of the eclipsing binary WOCS 11028 \citep{2021AJ....161...59S}, with one component near the cluster turnoff, 
provides an opportunity to test how well stellar models can describe the cluster. \citet{2021AJ....161...59S} found discrepancies of at least 0.17 mag ($\sim 10\sigma$) between the absolute magnitude of the primary component of the binary system and predictions by various stellar evolution codes, or equivalently a lower than predicted mass by $\delta m=0.05 M_{\odot}$, even for isochrones as young as 3.2 Gyr. \citet{2022A&A...665A.126N} with their updated PARSEC isochrones report similar discrepancies. 

\begin{figure}
\includegraphics[width=\columnwidth]{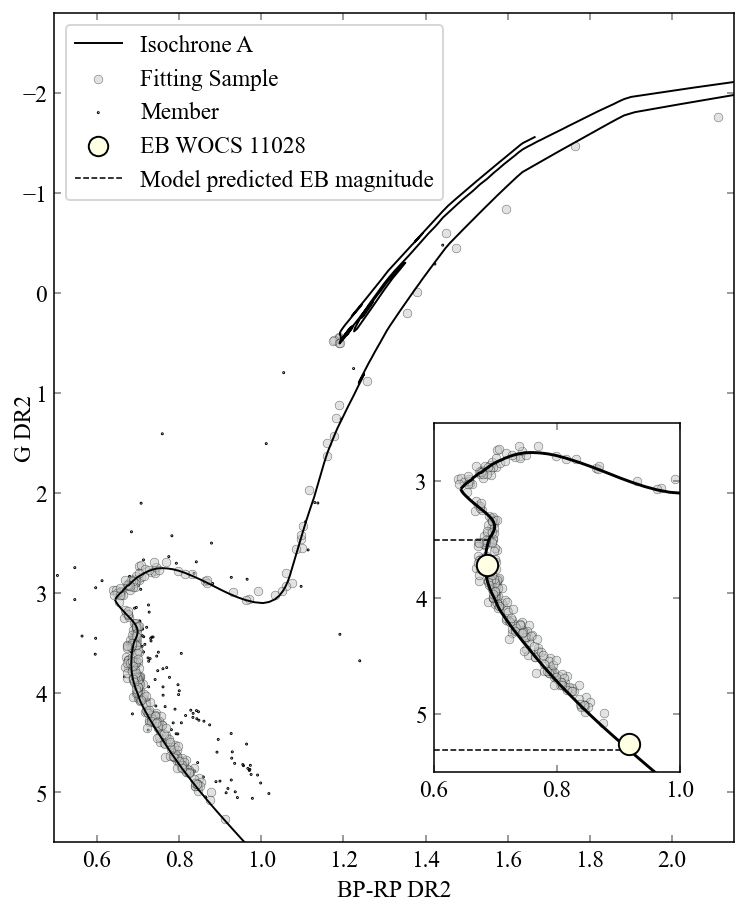}
\caption{The de-reddened colour-magnitude diagram of our M67 sample and the 3.95 Gyr old best fitting isochrone from this work, Isochrone A. In small black symbols: stars with RUWE>1.2, stars in multiple star systems, and stellar products of non-standard evolution, as well as suspected binary members manually removed from the sample (end of Section \ref{sec:cmd}). Several blue straggler members are bluer than 0.5 and are not shown in the figure. In grey symbols: presumed single members.  
Inset: the eclipsing binary system WOCS 11028 in large off-white circles resulting from the deconvolution of the binary's photometry, see Table \ref{tab:ebinary}. Dashed lines show the isochrone's predicted magnitudes for the masses of the two binary components. $\Delta G_{\mathrm{mag}}$ is $-0.05 \pm 0.09$, or $-0.67\sigma$ for the secondary component and $+0.21 \pm$ 0.06, or $+12\sigma$ for the primary component,  where $\sigma$ values correspond to those shown in Table~\ref{tab:ebinary}.}
\label{fig:isochrone}
\end{figure}

\begin{table*}
    \centering
    \begin{tabular}{cccccc}
Binary Member	&  BP-RP$^{*}$ &	$G^{*}$ & 2$\sigma$(G)	&Mass \msol\ &	Radius \rsol\ \\
    \hline
WOCS 11028a	 & 0.686  &13.336 & 0.035 & 1.222 $\pm$ 0.006&	1.430 $\pm$ 0.030\\
WOCS 11028b	& 0.916	 &14.871  & 0.150 & 0.909 $\pm$ 0.004&	0.904 $\pm$ 0.015\\
 \\
    \end{tabular}
    \caption{The eclipsing binary results from \citet{2021AJ....161...59S}. $^{*}$BP-RP and G are from their main sequence fit in Gaia DR2 magnitudes and have been dereddened and corrected for extinction using $A_{V}$ from dust maps.  }
    \label{tab:ebinary}
\end{table*}

Our Isochrone A-predicted G-magnitudes for the eclipsing primary and secondary (Figure~\ref{fig:isochrone}, inset) are respectively, $+0.21 \pm 0.06$ and $-0.05 \pm 0.09$ magnitudes relative to the eclipsing binary estimates in Table~\ref{tab:ebinary}. This includes uncertainties from our distance modulus and from the magnitudes and masses of the eclipsing binary. These offsets are comparable to the ones obtained by \citet{2021AJ....161...59S} and \citet{2022A&A...665A.126N}. Like us, \citet{2022A&A...665A.126N} found isochrones that agreed with the secondary within uncertainties but not with the primary. This is concerning because \citet{2021AJ....161...59S} reported lower confidence in estimating the secondary star's magnitude compared to the primary.

\subsubsection{Alternative Isochrone EB}

\begin{figure}
\centering
\includegraphics[width=\columnwidth]{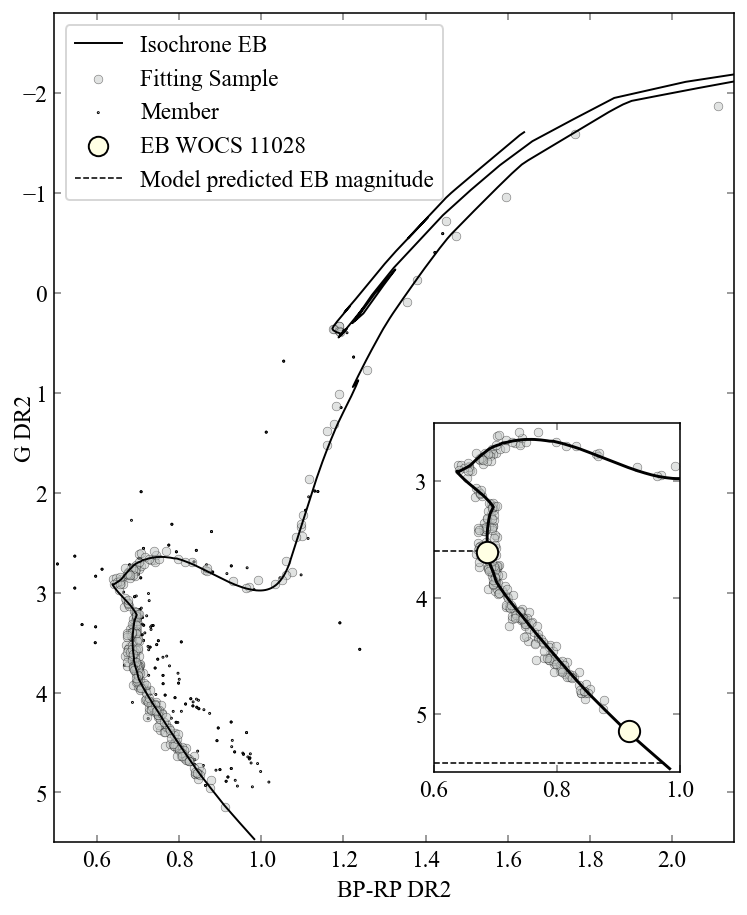}
  \caption{Alternative Isochrone EB. Dashed lines in the inset show predicted magnitudes for the masses of the two binary components. $\Delta G_{\mathrm{mag}}$ is $0.01 \pm 0.06$ ($0.3\sigma$) for the primary component and $-0.27 \pm0.09$, or $-4\sigma$ for the secondary component,  where $\sigma$ values correspond to those shown in Table~\ref{tab:ebinary}. Symbols follow Figure\ref{fig:isochrone}. }
\label{fig:isoeb}
\end{figure}

\begin{figure}
\centering
\includegraphics[width=\columnwidth]{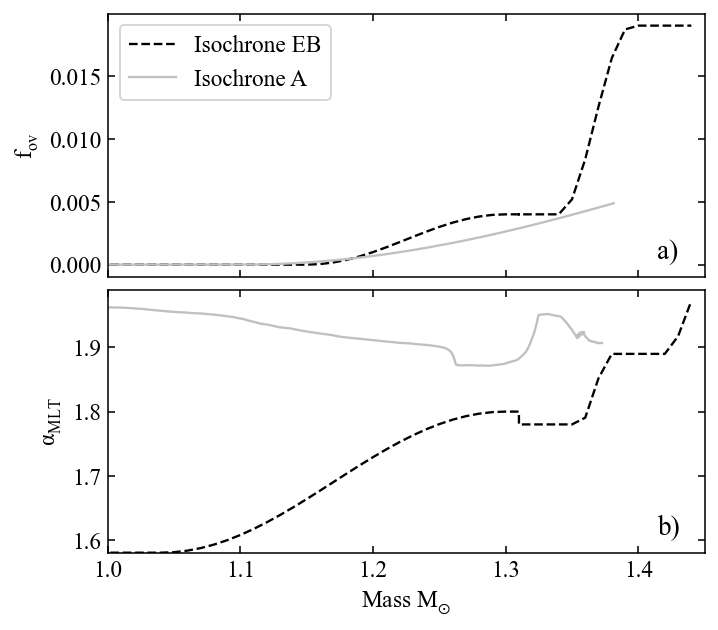}
\caption{(a) The overshooting parameter $f_{\mathrm{ov}}$ and the mixing length parameter \amlt\ used in Isochrone EB (dashed) shown along the values used in Isochrone A (solid grey). }
\label{fig:isoeb2}
\end{figure}

To address the eclipsing binary issue, we attempted to anchor an alternative isochrone on the mass and magnitude of the primary (WOCS 11028a) within $1\sigma$, as shown in the inset of Figure~\ref{fig:isoeb}.
To match the mass of the primary, we needed an isochrone 
with a main-sequence absolute brightness that is 0.2 magnitudes fainter compared to Isochrone A for a given mass.
This can be achieved by making specific parameter adjustments, such as lowering the helium content, changing the reference element mixture, reducing isochrone age, increasing the distance modulus, or adding more reddening. Although all modifications will also move the secondary, applying one or more of these modifications could reconcile discrepancies with the primary, but it may also result in inconsistencies in other parts of the isochrone. For instance, a lower isochrone age leads to a subgiant branch morphology that does not match the M67 observed data, but that, in turn, could be resolved with increased shell overshoot values. 
Considering the above, we developed a few alternative isochrones that could be anchored on WOCS 11028a. None of them, however, could match both binary components simultaneously, and all of them violated at least one other constraint. 
We consider that some constraints are more reliably established than others. The cluster's reddening has been consistently measured at $E(B-V)=0.04 \pm 0.04$ magnitudes, or its equivalent in other bands \citep{1987AJ.....93..634N, 1998ApJ...500..525S, 2007AJ....133..370T}, which agrees with our calculations from \citet{2016A&A...594A..13P} dust maps. Additionally, the helium content should not be much lower than the Sun's, given the near-solar age and metallicity of the cluster, and it should certainly be higher than the primordial helium abundance. In contrast, historical discrepancies in the parallax measurements for M67 exist (see Section \ref{sec:distance}), and convective overshoot is known to be a significant source of uncertainty in stellar models \citep{2013sse..book.....K,  2017RSOS....470192S, 2020ApJ...904...22V}. Therefore, we present here our best-fit alternative model, Isochrone EB, which varies these less-established constraints.
We achieve a good match between Isochrone EB and the data by using: (a) the solar mixture from \citet{1998SSRv...85..161G} with initial $Z=0.0196$, $Y=0.2710$ ($\Delta Y/\Delta Z=1.18$), (b) a distance modulus of 9.73, $2.4\sigma$ larger than our derived value from zero-point corrected Gaia DR3 parallaxes, (c) and a sharp and significant increase in overshoot at and beyond $1.32\,\mathrm{M}_{\odot}$, with manually adjusted values of \amlt.
The model parameters of Isochrone EB are summarised in Table~\ref{tab:isoeb}, and a comparison of its overshoot and mixing length to those of Isocrone A are shown in Figures~\ref{fig:isoeb2}a and b, respectively.
\begin{table}
    \centering
    \begin{tabular}{ll}
        \hline
        Parameter & Isochrone EB\\
        \hline
        Age & 3.55 Gyrs\\
        Distance modulus & 9.73\\
        Nuclear reaction network & h1, he3, he4, c12, n14, o16, ne20, mg24\\
        & (MESA's basic.net)\\
        Solar mixture & GS98 \citep{1998SSRv...85..161G}\\
        Z & 0.0196\\
        Y & 0.2710\\
        $\Delta$Z/$\Delta$Y & 1.18\\
        EoS &   HELM \citep{2000ApJS..126..501T}\\
         & + Skye \citep{2021ApJ...913...72J}\\
          & +  FreeEOS \citep{FreeEOS}\\
          & + OPAL \citep{2002ApJ...576.1064R}\\
          & + SCVH \citep{1995ApJS...99..713S}\\
        Interior opacities & OPAL \citep{1993ApJ...412..752I, 1996ApJ...464..943I}\\
        Low temperature opacities & FA05 \citep{2005ApJ...623..585F} \\
        Atmospheres & $T(\tau)$, varying, Eddington\\
        Overshooting & Exponential, increasing with mass,\\
        & sharp increase at M=1.32\msol\\
        \amlt\ & Variable, adjusted manually \\
        MLT\_option & Henyey\\
        Mass loss & No mass loss\\
        Convection & Schwarzschild criterion\\
        Diffusion & Yes\\
        \hline
    \end{tabular}
    \caption{Summary of parameters in MESA models for Isochrone EB shown in Figure\ref{fig:isoeb}.}
    \label{tab:isoeb}
\end{table}

We note that in Isochrone EB (Figure~\ref{fig:isoeb2}a), the values of overshooting in stars more massive than 1.35\msol\ make the models exceed the upper limit given by the Roxburgh criterion for small convective cores, where overshooting is limited to a level that would prevent the core from growing to more than 18\% of its `no-overshoot' size \citep{1992A&A...266..291R}. Furthermore, the study of convective overshooting remains a very active field, with various stellar signatures being used to calibrate models in different studies. Still, all recent studies find either a gradual increase with mass \citep{2024arXiv240212461L} or no increase at all \citep{2018A&A...618A.177C}.

From Figure~\ref{fig:isoeb2}b, we see some correspondence between the shapes described by both \amlt\ curves if we consider that the masses of Isochrone EB are close to $0.5\, $\msol\ larger than Isochrone A's at any given point in the colour-magnitude or $\log(g)$-\teff\ diagrams. However, the \amlt\ values we applied to stellar models of different masses are not based on any known physics but are solely adapted by eye to fit the CMD diagram.
From all the above, we conclude that even though Isochrone EB matches the mass and magnitude of WOCS11028a, the isochrone is unlikely to be a good model for the cluster. Also, preliminary results indicate that the luminosities of the giants predicted by the alternative isochrone EB are in tension with asteroseismic observations (Reyes et al., in preparation). 

\subsubsection{Possible causes of the model-mass conflict}

In addition to potential remaining shortcomings in stellar modelling, possible explanations for the conflict with the eclipsing binary masses include non-standard evolution of at least one component of WOCS11028, or systematics in the estimations of the eclipsing binary parameters. 
However, the presence of lithium in the spectra of the binary makes it unlikely that these stars have interacted in the past or resulted from a merger event, as, in such scenarios, lithium would have been consumed by the original lower mass stars with deeper surface convection zones, before merging \citep{2021AJ....161...59S}.
Possible grounds for an altered eclipsing binary mass or luminosity would be the non-detected presence of a third star in the WOCS 11028 system. Such has been the case in the past in the young Hyades open cluster and the multiple system HD27130 where a previous mass overestimate due to an unseen third
component led \citet{2001A&A...374..540L} to estimate a helium content of $Y=0.255$, while more recent studies indicate a higher than solar helium abundance for the Hyades \citep[see][and references therein]{2023MNRAS.518..662B}. For a discussion on the effects of HD27130 on the derived properties of the Hyades see \citet{2021A&A...645A..25B}.
However, even if there were a faint third star hiding in the photometry of WOCS11028 due to a particularly pathological setup of the multiple system, it would mean a similar and small mass shift to both known components, with no significant change in their relative positions in the colour-magnitude diagram. This opens up the possibility that models do not accurately represent the speed at which stars evolve during the main sequence, at least for stars 
around $1.2 \mathrm{M}_{\odot}$, which could make us reconsider our understanding of stellar evolution. In summary, the case of WOCS11028 is still very much open and requires further investigation.

\section{Age uncertainty through chi-square analysis}
\label{sec:chisq}

Although we followed the traditional qualitative approach of matching the input physics of Isochrone A to the M67 colour-magnitude diagram, we will in the following look to estimate the age uncertainty using a $\chi^{2}$ approach. We take our isochrone-fitting sample from section \ref{sec:cmd} and further remove the stars that have evolved beyond the point where the red giant branch leaves the $\log(g)$-\teff\ area covered by the 3D grid, marked by the segmented line in Figure~\ref{fig:isocolor}b. 

\begin{figure}
\centering
\includegraphics[width=\columnwidth]{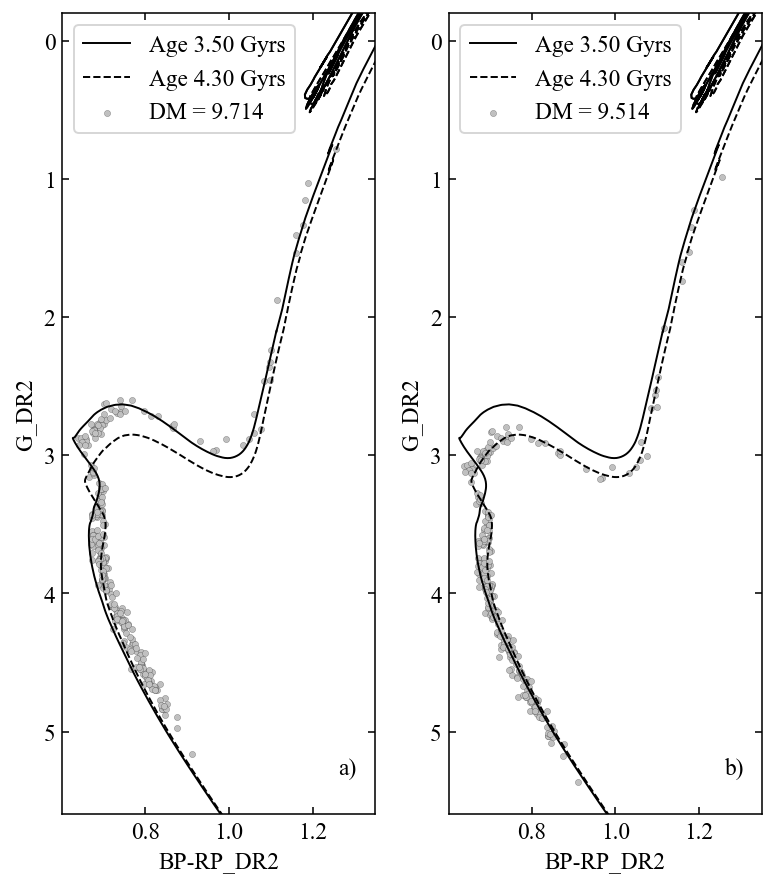}
\caption{Isochrones of 3.50 Gyrs (solid) and 4.3 Gyrs (dashed) using the same physics as Isochrone A. (a) M67 isochrone fitting sample with a distance modulus is 9.714. (b) same as panel a, but for a distance modulus of 9.514.}
\label{fig:agedm}
\end{figure}

We vary the model age between 3.5 and 4.3 Gyrs, covering the age range quoted in literature \citep{2004PASP..116..997V, 2009ApJ...698.1872S, 2016ApJ...823...16B}, and the distance modulus within 2$\sigma$ of our Gaia DR3 parallax-derived value of $9.614 \pm 0.049$. Figure~\ref{fig:agedm} shows the extremes of varying age and distance modulus like this.

\begin{figure*}
\centering
\includegraphics[width=\textwidth]{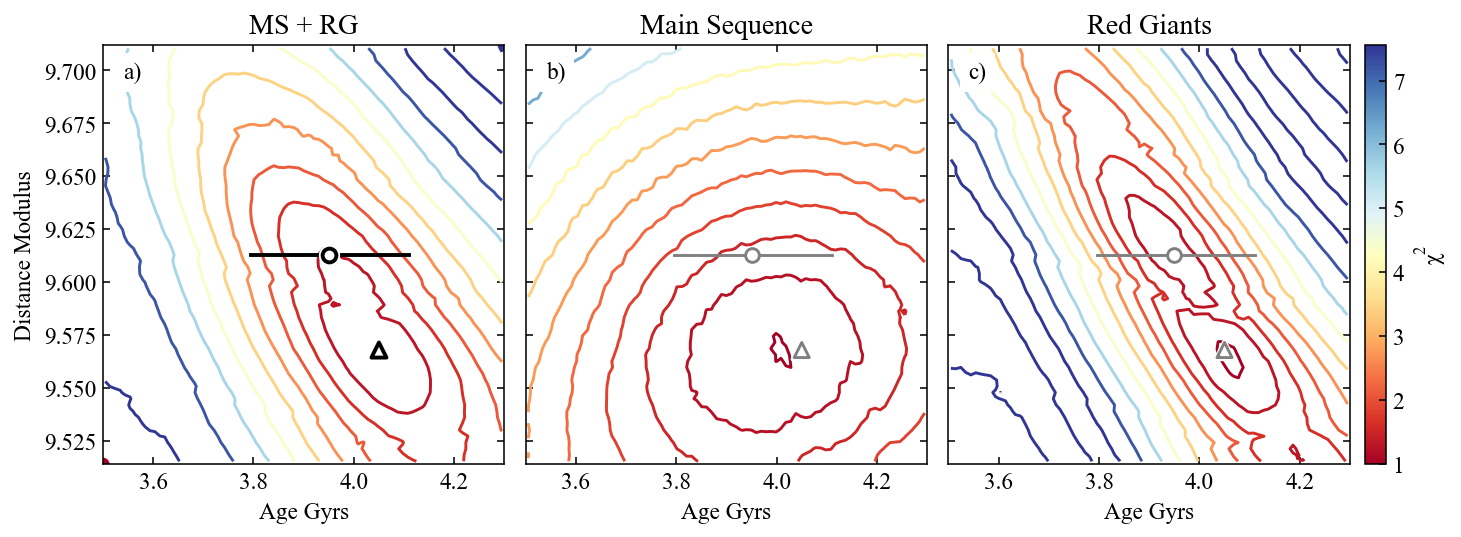}
\caption{$\chi^{2}$-like statistics from M67 main sequence and giant stars for isochrone ages between 3.5 and 4.3 Gyrs and distance modulus within 2$\sigma$ of our Gaia parallax-derived value of 9.614 $\pm$ 0.049. The triangle shows the overall lowest $\chi^{2}$ for panel a, and the circle, including errorbars, shows our quoted age of $3.95^{+ 0.16}_{- 0.15}$ Gyrs, corresponding to the lowest $\chi^{2}$ at distance modulus = 9.614. The position of the circle and the triangle are copied in the other panels for completeness.}
\label{fig:chisquare}
\end{figure*}

For each combination of age and distance modulus, we used the nearest matching point on the isochrone as the expected value ($\vec{x}_{\mathrm{exp},i}$) for each star ($\vec{x}_{\mathrm{obs},i}$) in a $\chi^{2}$-like goodness of fit analysis. Following \citet{2021AJ....161...59S}, we define a $\chi^{2}$-like parameter involving fractional colour and magnitude differences, of the form:
\begin{align}
    \frac{1}{N} \sum_{i=0}^{N} \frac{(\vec{x}_{\mathrm{obs},i} - \vec{x}_{\mathrm{exp},i})^{2}}{\vec{x}_{\mathrm{exp},i}^{2}},  \;  \; \vec{x}=(\mathrm{col}_{i}, \mathrm{mag}_{i})
\end{align}
\noindent that we later rescale so that the lowest $\chi^{2}$ value equals 1. We do not include a $\sigma$ in this expression because Gaia does not provide magnitude uncertainties. 
Figure~\ref{fig:chisquare} shows the age versus distance modulus space colour-coded by this (rescaled) parameter.
Figure~\ref{fig:chisquare}a presents the combined statistics of both main-sequence and giant stars, which we will use in the following analysis.
We find that the lowest $\chi^{2}$ (black triangle) is at a distance modulus about 1$\sigma$ below the parallax-based value (black circle), and with a corresponding age of $4.04\, \mathrm{Gyrs}$. If this were to be caused by any systematics in Gaia DR3 parallaxes, it would correspond to about $25\,\mu\mathrm{as}$. However, there is no indication that such systematics exist beyond the \citet{2021A&A...649A...4L} corrections \citep{2021AJ....161..214Z}, which we have already included. Furthermore, a smaller distance modulus would increase discrepancies between models and WOCS 11028. We, therefore, prefer to quote the age as $3.95\,\mathrm{Gyrs}$ --the black circle in Figure~\ref{fig:chisquare}a -- and our confidence interval as $[3.80, 4.11]\, \mathrm{Gyrs}$, the point where the $\chi^{2}$ has risen by 1 from its minimum value along the adopted parallax-based distance modulus (black error bars).
For completeness, we also present the $\chi^{2}$ results restricted to main sequence stars in Figure~\ref{fig:chisquare}b, and giants in Figure~\ref{fig:chisquare}c, where the split into those groups is defined by core hydrogen exhaustion (the hottest point along the isochrone). For reference, we copy in the latter panels with grey symbols the location of the lowest $\chi^{2}$ points from panel a. 

\section{Conclusions and Future Work}
We have presented a detailed isochrone, Isochrone A, specifically designed for the low-reddening cluster M67 using state-of-the-art stellar input physics. Our models match the morphology of the tight cluster sequence better than previously reported in the literature in Gaia colours and magnitudes. 
Our results indicate that adopting a mixing length \amlt\ that is dependent on \teff\ and $\log(g)$, such as the one obtained from the 3D-grid, can improve the fit of models to observations. Therefore, this approach should be considered in future standard stellar models.
Our best age estimate is $3.95^{+ 0.16}_{- 0.15}\,\mathrm{Gyrs}$, in agreement with the estimates from \citet{2004PASP..116..997V} ($4.0 \pm 0.4\, \mathrm{Gyrs}$ from $BV$ photometry), \citet{2009ApJ...698.1872S} ($3.5-4.0\, \mathrm{Gyrs}$ from 2MASS photometry), \citep{2016ApJ...823...16B} ($4.2 \pm 0.2\, \mathrm{Gyrs}$ from gyrochronology), \citet{2012MNRAS.427..127B} ($3.7\, \mathrm{Gyrs}$ from $BV$ photometry using PARSEC isochrones v1.2), and \citet{2022A&A...665A.126N} ($3.9\, \mathrm{Gyrs}$ from Gaia DR2 photometry using PARSEC isochrones v2.0).

Despite the improvements in the match to the cluster colour-magnitude diagram, our isochrone shows some disagreement with the measured mass for the primary component of the eclipsing binary WOCS11028 near the cluster turnoff. The only alternative isochrone we could anchor on the primary eclipsing binary is based on improbable input physics and fundamental cluster properties.
Therefore, we conclude that the literature values of the masses and luminosities for the eclipsing binary system are incompatible with the rest of the cluster in a scenario where both binary components have evolved independently. Of course, this is under the perhaps unrealistic assumption that there are no shortcomings in our current model input physics.

To help resolve the issues surrounding WOCS 11028, one could potentially use detailed asteroseismic modelling to confirm masses and radii of the cluster stars. In addition, high-resolution imaging of the binary might be able to detect any undiscovered component that could affect the mass and radius estimates.

\section*{Software}
Extinction correction performed using the \textit{dustmaps} python tool \citep{2018JOSS....3..695M}. \url{https://dustmaps.readthedocs.io/en/latest/}

Interpolation to bolometric correction tables was performed using the python package \textit{isochrones} v2.1 \citep{2015ascl.soft03010M}. 
\url{https://isochrones.readthedocs.io/en/latest/}

Isochrones are constructed using the fortran code \textit{iso}  described in \citet{2016ascl.soft01021D, 2016ApJS..222....8D}. 
\url{https://cmasher.readthedocs.io/user/introduction.html}

\section*{Acknowledgements}

D.S. is supported by the Australian Research Council (DP190100666).
This work has made use of data from the European Space Agency (ESA) mission
{\it Gaia} (\url{https://www.cosmos.esa.int/Gaia}), processed by the {\it Gaia}
Data Processing and Analysis Consortium (DPAC,
\url{https://www.cosmos.esa.int/web/Gaia/dpac/consortium}). Funding for the DPAC
has been provided by national institutions, in particular the institutions
participating in the {\it Gaia} Multilateral Agreement.

This research includes computations using the computational cluster Katana supported by Research Technology Services at UNSW Sydney, \url{https://doi.org/10.26190/669x-a286}.

We would like to thank the participants of the MIAPbP Program "Stellar Astrophysics" in Garching August 2023 for the discussion on artificial metallicity trends with $\log(g)$ when not using 3D-NLTE spectroscopic analyses.

\section*{Data Availability}
 \label{sec:Data}
The Isochrone and the inlists for the models presented in this work can be found on the Zenodo online software platform, \url{https://zenodo.org/records/12616441}.

\appendix

\section{Isochrone A}
Our final 3.95 Gyr isochrone, Isochrone A, is partially presented in \autoref{tab:isocsv}, and the full isochrone from main sequence to core helium burning is available for download through the online version of the paper. The column names correspond to standard MESA parameters, except mixing\_length\_alpha and f\_overshoot, which correspond to \amlt\ and $f_{\mathrm{ov}}$, respectively, as denoted in the text.
\begin{table*}
    \centering
    \begin{tabular}{ccccccccc}
    isochrone\_age\_yr & star\_mass  & log\_L & log\_Teff & log\_R & log\_g & ... & mixing\_length\_alpha & f\_overshoot  \\
\hline
3.9500e9 &0.8506 &-0.4003 &3.7133 &-0.1038 &4.5755& ... &1.9849 &0.0000 \\
3.9500e9 &0.8591 &-0.3778 &3.7165 &-0.0989 &4.5700& ... &1.9774 &0.0000 \\
3.9500e9 &0.8676 &-0.3557 &3.7195 &-0.0940 &4.5644& ... &1.9691 &0.0000 \\
3.9500e9 &0.8759 &-0.3339 &3.7226 &-0.0891 &4.5587& ... &1.9637 &0.0000 \\
3.9500e9 &0.8841 &-0.3127 &3.7255 &-0.0843 &4.5532& ... &1.9602 &0.0000 \\
    \end{tabular}
    \caption{First few rows of selected columns of Isochrone A. The full isochrone, including key internal structure parameters and photometric bands, is available for download from the online version of the paper, or on Zenodo (See Data Availability.)}
    \label{tab:isocsv}
\end{table*}

\section{Other Isochrones}
\label{sec:other_isos}
\begin{figure*}
\includegraphics[width=\textwidth]{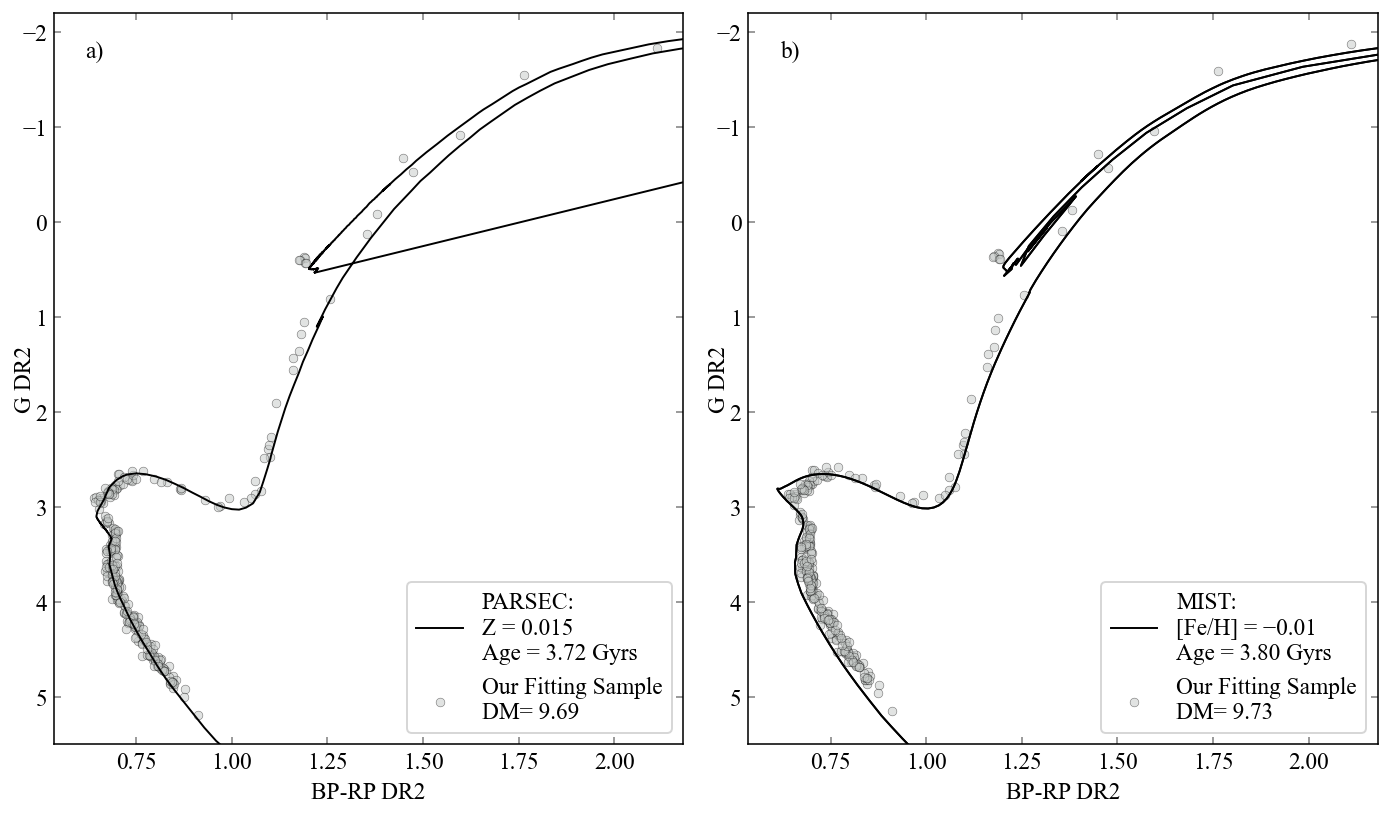}
\caption{Best fit isochrones to the M67 Gaia colour magnitude diagram from a) \citet{2022A&A...665A.126N}, and b) \citet{2018ApJ...863...65C}.}
\label{fig:other_isos}
\end{figure*}

Figure~\ref{fig:other_isos} shows isochrones from two recent works that have been fitted to the M67 colour-magnitude diagram in Gaia colours. These are models designed for general use and, therefore, not specifically for M67. 
\citet{2022A&A...665A.126N} use The PAdova and tRieste Stellar Evolutionary Code v2.0, and similarly to Isochrone A from this work, incorporates a mass-dependent overshoot and the AESOPUS opacity tables. Their best fit to M67 is shown with our stellar fitting sample in Figure~\ref{fig:other_isos}a. 
\citet{2018ApJ...863...65C} -the MIST models- use the MESA evolutionary code, and a fixed core overshoot $f_{\mathrm{ov}} = 0.016$. Figure~\ref{fig:other_isos}b shows their fit to M67. In each panel, the fitting sample is corrected to absolute magnitudes using the distance modulus adopted by the corresponding authors.

\begin{figure*}
\begin{minipage}[b]{\textwidth}
\centering
\includegraphics[width=\textwidth]{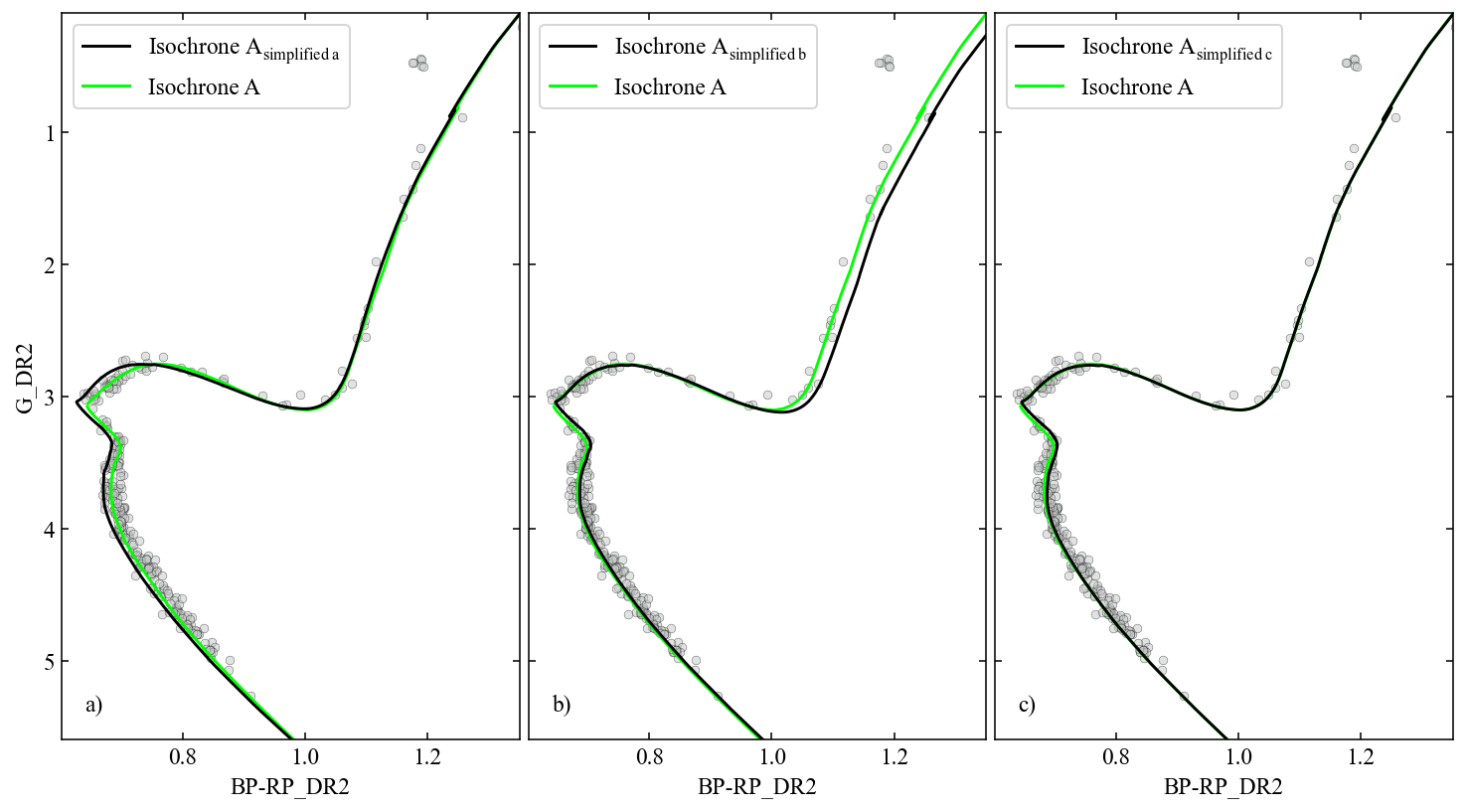}
\end{minipage}

\vspace{5pt}
\begin{minipage}[b]{\textwidth}
    \centering
    \begin{tabular}{l|cccc}
         \textbf{Parameter} &\textbf{(a)}&  \textbf{(b)}&  \textbf{(c)}& \textbf{Isochrone A}\\
         \hline
         \textbf{$\alpha_{\mathrm{MLT}}$} &  1.91& 3D grid & 3D grid& 3D grid\\
         \textbf{T$(\tau)$} &  Eddington& Trampedach solar & Trampedach solar & Trampedach solar\\
         \vspace{3pt}
         \textbf{photospheric transition}&  default& improved & improved & improved\\
         \vspace{3pt}
         \textbf{low T opacities}& FA05 & FA05 & AESOPUS& AESOPUS\\
         \vspace{3pt}
         \textbf{core overshoot}& fixed $f_{\mathrm{ov}=0.004}$ & fixed $f_{\mathrm{ov}=0.004}$ & fixed $f_{\mathrm{ov}=0.004}$& mass-dependent\\
    \end{tabular}
\end{minipage}
\vspace{5pt}   

\caption{In b) and c), the isochrones in black show the effects of cumulatively applying parameter `improvements' to the simplified form of Isochrone A shown in (a). The table below the figure shows the model parameters we vary and their values, along with the values used in Isochrone A.}
\label{fig:iso_other}
\end{figure*}

Figure~\ref{fig:iso_other} shows the qualitative effects of our atmosphere and core-overshoot choices from sections \ref{sec:atmosphere} and \ref{sec:overshoot} in our models. 
In all panels, Isochrone A is shown in green and: Panel (a) shows a base-model from where we start: similar to the MIST models, but with our chosen initial composition (\ref{sec:feh}), age, and the carefully chosen (fixed) values of overshoot $f_{\mathrm{ov}}$ and \amlt\ to best fit M67; Column (a) in the table just below the figure indicates five key parameter values.
We see that the simplified model can still be made to fit the red giant branch well with \amlt=1.91, but the model is too hot compared with the main-sequence. 
In (b) we show the base model from (a), but now applying 
the 3D-grid-based varying \amlt\ and associated atmospheric improvements from \citet{2014MNRAS.442..805T, 2014MNRAS.445.4366T} (see column (b) in table). These changes move the isochrone towards colder temperatures, making the model fit the main sequence and subgiants better, however the red giant branch is now "too cold", at least with the age and [Fe/H] of Isochrone A.
In (c) we now also add low opacity tables that account for CNO processes in red giants (see column (c)). This brings the model back into agreement with the red giants. The only remaining simplified parameter in (c) is the fixed overshoot $f_{\mathrm{ov}}$, which has a subtle, but noticeable, effect on the morphology of the main sequence turnoff, where the even smaller values of $f_{\mathrm{ov}}$ used in Isochrone A at these masses (Figure~\ref{fig:isocolor}c, d) appear to fit the stellar sequence marginally better.

\section{Extinction Maps}
\label{sec:map}
\begin{figure*}
\includegraphics[width=0.75\textwidth]{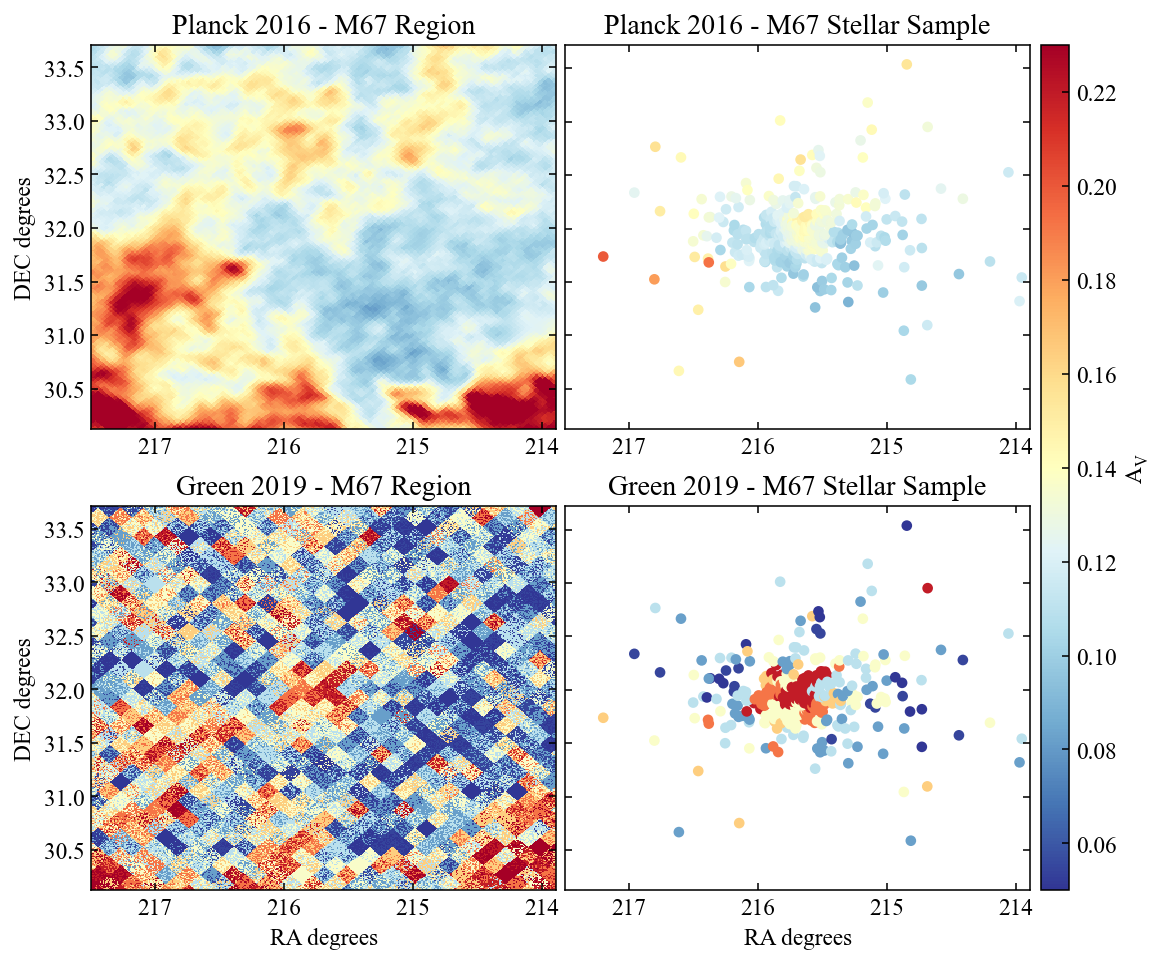}
\caption{Extinction maps in galactic coordinates of the M67 cluster region (left) and of our M67 stellar sample (right). The colour expresses the extinction $A_{V}$ at each point of the sky.}
\label{fig:maps}
\end{figure*}

We present extinction maps of the region around the M67 cluster, centred at $\mathrm{RA}=215.69^{\circ}$, $\mathrm{Dec}=+31.92^{\circ}$ (galactic coordinates), that we considered for differential de-reddening of M67 photometry, aiming to reduce the scatter of the cluster sequence in the colour-magnitude diagram.
We compared the 2D extinction map by \citet{2016A&A...594A..13P} with the Bayestar 3D reddening map \citep{2019ApJ...887...93G} and used the coefficients from Table 6 from \citet{Schlafly_2011} to convert from Bayestar reddening to $A_{V}$ extinctions.
We found that the 2D map results in a smoother map of extinctions that works well to reduce the scatter in the M67 colour-magnitude diagram (Figure~\ref{fig:reddening}), while the 3D Bayestar map does not behave well, at least in this area of the sky, and at the small scales we require. The 3D extinction map looks pixelated and shows sharp variations, as seen in Figure~\ref{fig:maps}. This could be related to the systematic trends in reddening identified by \citet{2019ApJ...887...93G} at low reddenings. This is a note of caution for future cluster studies.

\section{Metallicity trend in Red Giants from APOGEE DR17}
\label{sec:apogee}
\begin{figure*}
\includegraphics[width=\textwidth]{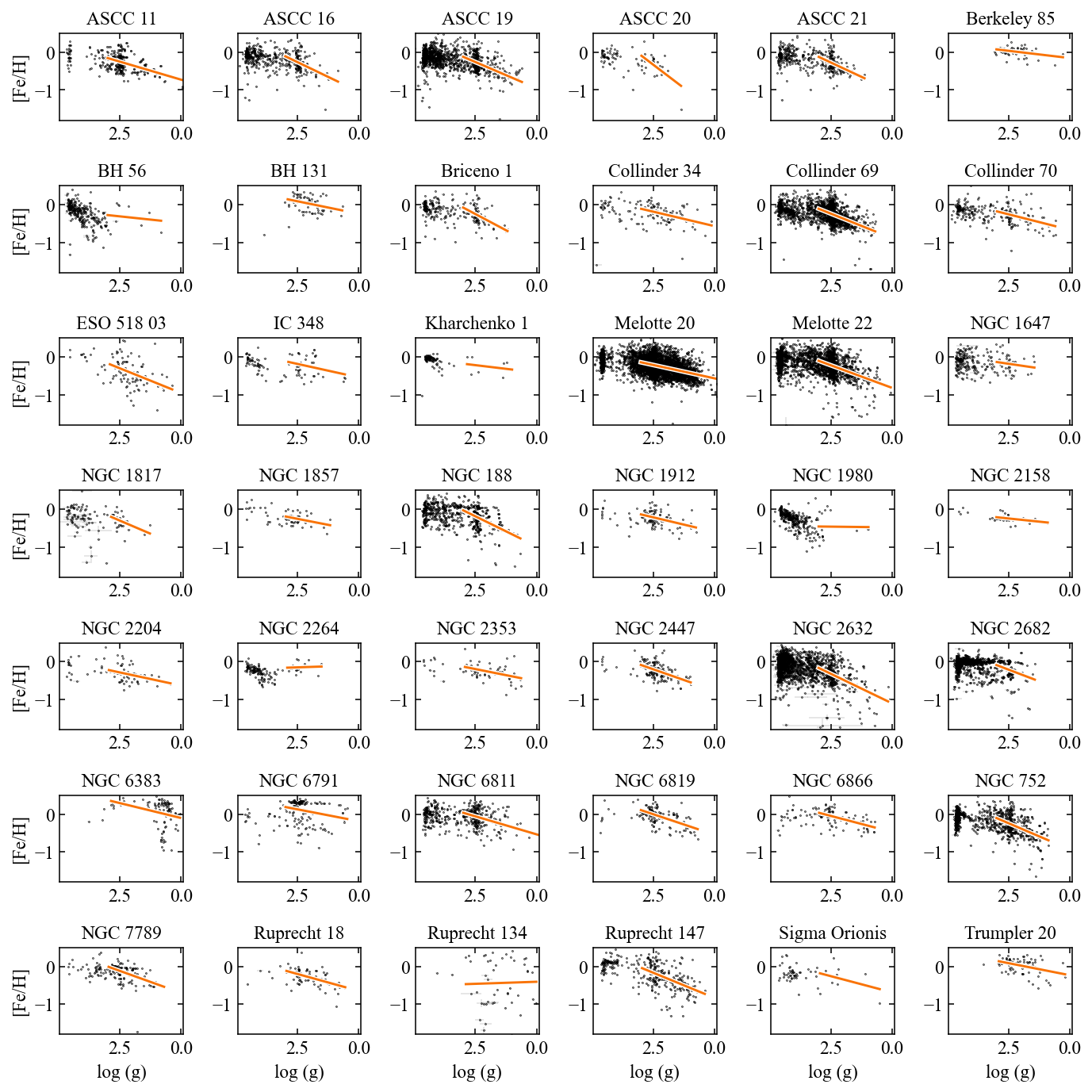}
\caption{Clusters from the APOGEE DR17 Open Cluster Chemical Abundances and Mapping Catalogue with more than 80 cluster members (alphabetically ordered). The orange lines show a fit to the metallicities of all stars with $\log(g)$<= 3.0 within each cluster.}
\label{fig:apogee1}
\end{figure*}

The trend of decreasing metallicity with decreasing $\log(g)$ observed in M67 was observed in almost all of the clusters included in the APOGREE DR17 Open Cluster Chemical Abundances and Mapping Catalogue \footnote{https://www.sdss4.org/dr17/data\_access/value-added-catalogs/?vac\_id=open-cluster-chemical-abundances-and-mapping-catalog}.
In Figure~\ref{fig:apogee1} we show the clusters in the catalogue with more than 80 members, and fit the metallicity trend for $\log (g)<3.0$ in each cluster.

\bibliographystyle{mnras}
\bibliography{M67} 

\begin{thebibliography}{}
\makeatletter
\relax
\def\mn@urlcharsother{\let\do\@makeother \do\$\do\&\do\#\do\^\do\_\do\%\do\~}
\def\mn@doi{\begingroup\mn@urlcharsother \@ifnextchar [ {\mn@doi@} {\mn@doi@[]}}
\def\mn@doi@[#1]#2{\def\@tempa{#1}\ifx\@tempa\@empty \href {http://dx.doi.org/#2} {doi:#2}\else \href {http://dx.doi.org/#2} {#1}\fi \endgroup}
\def\mn@eprint#1#2{\mn@eprint@#1:#2::\@nil}
\def\mn@eprint@arXiv#1{\href {http://arxiv.org/abs/#1} {{\tt arXiv:#1}}}
\def\mn@eprint@dblp#1{\href {http://dblp.uni-trier.de/rec/bibtex/#1.xml} {dblp:#1}}
\def\mn@eprint@#1:#2:#3:#4\@nil{\def\@tempa {#1}\def\@tempb {#2}\def\@tempc {#3}\ifx \@tempc \@empty \let \@tempc \@tempb \let \@tempb \@tempa \fi \ifx \@tempb \@empty \def\@tempb {arXiv}\fi \@ifundefined {mn@eprint@\@tempb}{\@tempb:\@tempc}{\expandafter \expandafter \csname mn@eprint@\@tempb\endcsname \expandafter{\@tempc}}}

\bibitem[\protect\citeauthoryear{{Alam} et~al.,}{{Alam} et~al.}{2015}]{2015ApJS..219...12A}
{Alam} S.,  et~al., 2015, \mn@doi [\apjs] {10.1088/0067-0049/219/1/12}, \href {https://ui.adsabs.harvard.edu/abs/2015ApJS..219...12A} {219, 12}

\bibitem[\protect\citeauthoryear{{Asplund}, {Grevesse}, {Guedel}  \& {Sauval}}{{Asplund} et~al.}{2005}]{2005astro.ph.10377A}
{Asplund} M.,  {Grevesse} N.,  {Guedel} M.,   {Sauval} A.~J.,  2005, \mn@doi [arXiv e-prints] {10.48550/arXiv.astro-ph/0510377}, \href {https://ui.adsabs.harvard.edu/abs/2005astro.ph.10377A} {pp astro--ph/0510377}

\bibitem[\protect\citeauthoryear{{Asplund}, {Grevesse}, {Sauval}  \& {Scott}}{{Asplund} et~al.}{2009}]{2009ARA&A..47..481A}
{Asplund} M.,  {Grevesse} N.,  {Sauval} A.~J.,   {Scott} P.,  2009, \mn@doi [\araa] {10.1146/annurev.astro.46.060407.145222}, \href {https://ui.adsabs.harvard.edu/abs/2009ARA&A..47..481A} {47, 481}

\bibitem[\protect\citeauthoryear{{Ball}}{{Ball}}{2021}]{2021RNAAS...5....7B}
{Ball} W.~H.,  2021, \mn@doi [Research Notes of the American Astronomical Society] {10.3847/2515-5172/abd9cb}, \href {https://ui.adsabs.harvard.edu/abs/2021RNAAS...5....7B} {5, 7}

\bibitem[\protect\citeauthoryear{{Barker} \& {Paust}}{{Barker} \& {Paust}}{2018}]{2018PASP..130c4204B}
{Barker} H.,  {Paust} N. E.~Q.,  2018, \mn@doi [\pasp] {10.1088/1538-3873/aaa597}, \href {https://ui.adsabs.harvard.edu/abs/2018PASP..130c4204B} {130, 034204}

\bibitem[\protect\citeauthoryear{{Barnes}, {Weingrill}, {Fritzewski}, {Strassmeier}  \& {Platais}}{{Barnes} et~al.}{2016}]{2016ApJ...823...16B}
{Barnes} S.~A.,  {Weingrill} J.,  {Fritzewski} D.,  {Strassmeier} K.~G.,   {Platais} I.,  2016, \mn@doi [\apj] {10.3847/0004-637X/823/1/16}, \href {https://ui.adsabs.harvard.edu/abs/2016ApJ...823...16B} {823, 16}

\bibitem[\protect\citeauthoryear{{Beeson} et~al.,}{{Beeson} et~al.}{2024}]{2024MNRAS.529.2483B}
{Beeson} K.~L.,  et~al., 2024, \mn@doi [\mnras] {10.1093/mnras/stae439}, \href {https://ui.adsabs.harvard.edu/abs/2024MNRAS.529.2483B} {529, 2483}

\bibitem[\protect\citeauthoryear{{Bertelli Motta} et~al.,}{{Bertelli Motta} et~al.}{2018}]{2018MNRAS.478..425B}
{Bertelli Motta} C.,  et~al., 2018, \mn@doi [\mnras] {10.1093/mnras/sty1011}, \href {https://ui.adsabs.harvard.edu/abs/2018MNRAS.478..425B} {478, 425}

\bibitem[\protect\citeauthoryear{{Brandner}, {Calissendorff}  \& {Kopytova}}{{Brandner} et~al.}{2023}]{2023MNRAS.518..662B}
{Brandner} W.,  {Calissendorff} P.,   {Kopytova} T.,  2023, \mn@doi [\mnras] {10.1093/mnras/stac2247}, \href {https://ui.adsabs.harvard.edu/abs/2023MNRAS.518..662B} {518, 662}

\bibitem[\protect\citeauthoryear{{Bressan}, {Marigo}, {Girardi}, {Salasnich}, {Dal Cero}, {Rubele}  \& {Nanni}}{{Bressan} et~al.}{2012}]{2012MNRAS.427..127B}
{Bressan} A.,  {Marigo} P.,  {Girardi} L.,  {Salasnich} B.,  {Dal Cero} C.,  {Rubele} S.,   {Nanni} A.,  2012, \mn@doi [\mnras] {10.1111/j.1365-2966.2012.21948.x}, \href {https://ui.adsabs.harvard.edu/abs/2012MNRAS.427..127B} {427, 127}

\bibitem[\protect\citeauthoryear{{Brogaard} et~al.,}{{Brogaard} et~al.}{2012}]{2012A&A...543A.106B}
{Brogaard} K.,  et~al., 2012, \mn@doi [\aap] {10.1051/0004-6361/201219196}, \href {https://ui.adsabs.harvard.edu/abs/2012A&A...543A.106B} {543, A106}

\bibitem[\protect\citeauthoryear{{Brogaard} et~al.,}{{Brogaard} et~al.}{2021}]{2021A&A...645A..25B}
{Brogaard} K.,  et~al., 2021, \mn@doi [\aap] {10.1051/0004-6361/202039250}, \href {https://ui.adsabs.harvard.edu/abs/2021A&A...645A..25B} {645, A25}

\bibitem[\protect\citeauthoryear{{Buder} et~al.,}{{Buder} et~al.}{2021}]{2021MNRAS.506..150B}
{Buder} S.,  et~al., 2021, \mn@doi [\mnras] {10.1093/mnras/stab1242}, \href {https://ui.adsabs.harvard.edu/abs/2021MNRAS.506..150B} {506, 150}

\bibitem[\protect\citeauthoryear{{Choi}, {Dotter}, {Conroy}, {Cantiello}, {Paxton}  \& {Johnson}}{{Choi} et~al.}{2016}]{2016ApJ...823..102C}
{Choi} J.,  {Dotter} A.,  {Conroy} C.,  {Cantiello} M.,  {Paxton} B.,   {Johnson} B.~D.,  2016, \mn@doi [\apj] {10.3847/0004-637X/823/2/102}, \href {https://ui.adsabs.harvard.edu/abs/2016ApJ...823..102C} {823, 102}

\bibitem[\protect\citeauthoryear{{Choi}, {Conroy}, {Ting}, {Cargile}, {Dotter}  \& {Johnson}}{{Choi} et~al.}{2018}]{2018ApJ...863...65C}
{Choi} J.,  {Conroy} C.,  {Ting} Y.-S.,  {Cargile} P.~A.,  {Dotter} A.,   {Johnson} B.~D.,  2018, \mn@doi [\apj] {10.3847/1538-4357/aad18c}, \href {https://ui.adsabs.harvard.edu/abs/2018ApJ...863...65C} {863, 65}

\bibitem[\protect\citeauthoryear{{Claret} \& {Torres}}{{Claret} \& {Torres}}{2017}]{2017ApJ...849...18C}
{Claret} A.,  {Torres} G.,  2017, \mn@doi [\apj] {10.3847/1538-4357/aa8770}, \href {https://ui.adsabs.harvard.edu/abs/2017ApJ...849...18C} {849, 18}

\bibitem[\protect\citeauthoryear{{Claret} \& {Torres}}{{Claret} \& {Torres}}{2018}]{2018ApJ...859..100C}
{Claret} A.,  {Torres} G.,  2018, \mn@doi [\apj] {10.3847/1538-4357/aabd35}, \href {https://ui.adsabs.harvard.edu/abs/2018ApJ...859..100C} {859, 100}

\bibitem[\protect\citeauthoryear{{Constantino} \& {Baraffe}}{{Constantino} \& {Baraffe}}{2018}]{2018A&A...618A.177C}
{Constantino} T.,  {Baraffe} I.,  2018, \mn@doi [\aap] {10.1051/0004-6361/201833568}, \href {https://ui.adsabs.harvard.edu/abs/2018A&A...618A.177C} {618, A177}

\bibitem[\protect\citeauthoryear{{Danielski}, {Babusiaux}, {Ruiz-Dern}, {Sartoretti}  \& {Arenou}}{{Danielski} et~al.}{2018}]{2018A&A...614A..19D}
{Danielski} C.,  {Babusiaux} C.,  {Ruiz-Dern} L.,  {Sartoretti} P.,   {Arenou} F.,  2018, \mn@doi [\aap] {10.1051/0004-6361/201732327}, \href {https://ui.adsabs.harvard.edu/abs/2018A&A...614A..19D} {614, A19}

\bibitem[\protect\citeauthoryear{{Dotter}}{{Dotter}}{2016a}]{2016ascl.soft01021D}
{Dotter} A.,  2016a, {ISO: Isochrone construction}, Astrophysics Source Code Library, record ascl:1601.021 (\mn@eprint {ascl} {1601.021})

\bibitem[\protect\citeauthoryear{{Dotter}}{{Dotter}}{2016b}]{2016ApJS..222....8D}
{Dotter} A.,  2016b, \mn@doi [\apjs] {10.3847/0067-0049/222/1/8}, \href {https://ui.adsabs.harvard.edu/abs/2016ApJS..222....8D} {222, 8}

\bibitem[\protect\citeauthoryear{{Ferguson}, {Alexander}, {Allard}, {Barman}, {Bodnarik}, {Hauschildt}, {Heffner-Wong}  \& {Tamanai}}{{Ferguson} et~al.}{2005}]{2005ApJ...623..585F}
{Ferguson} J.~W.,  {Alexander} D.~R.,  {Allard} F.,  {Barman} T.,  {Bodnarik} J.~G.,  {Hauschildt} P.~H.,  {Heffner-Wong} A.,   {Tamanai} A.,  2005, \mn@doi [\apj] {10.1086/428642}, \href {https://ui.adsabs.harvard.edu/abs/2005ApJ...623..585F} {623, 585}

\bibitem[\protect\citeauthoryear{{Gaia Collaboration}}{{Gaia Collaboration}}{2022}]{2022yCat.1355....0G}
{Gaia Collaboration} 2022, VizieR Online Data Catalog, \href {https://ui.adsabs.harvard.edu/abs/2022yCat.1355....0G} {p. I/355}

\bibitem[\protect\citeauthoryear{{Gaia Collaboration} et~al.,}{{Gaia Collaboration} et~al.}{2016}]{2016A&A...595A...1G}
{Gaia Collaboration} et~al., 2016, \mn@doi [\aap] {10.1051/0004-6361/201629272}, \href {https://ui.adsabs.harvard.edu/abs/2016A&A...595A...1G} {595, A1}

\bibitem[\protect\citeauthoryear{{Gaia Collaboration} et~al.,}{{Gaia Collaboration} et~al.}{2018a}]{2018A&A...616A...1G}
{Gaia Collaboration} et~al., 2018a, \mn@doi [\aap] {10.1051/0004-6361/201833051}, \href {https://ui.adsabs.harvard.edu/abs/2018A&A...616A...1G} {616, A1}

\bibitem[\protect\citeauthoryear{{Gaia Collaboration} et~al.,}{{Gaia Collaboration} et~al.}{2018b}]{2018A&A...616A..10G}
{Gaia Collaboration} et~al., 2018b, \mn@doi [\aap] {10.1051/0004-6361/201832843}, \href {https://ui.adsabs.harvard.edu/abs/2018A&A...616A..10G} {616, A10}

\bibitem[\protect\citeauthoryear{{Gaia Collaboration} et~al.,}{{Gaia Collaboration} et~al.}{2021}]{2021A&A...649A...1G}
{Gaia Collaboration} et~al., 2021, \mn@doi [\aap] {10.1051/0004-6361/202039657}, \href {https://ui.adsabs.harvard.edu/abs/2021A&A...649A...1G} {649, A1}

\bibitem[\protect\citeauthoryear{{Gaia Collaboration} et~al.,}{{Gaia Collaboration} et~al.}{2022}]{2022arXiv220605595G}
{Gaia Collaboration} et~al., 2022, \mn@doi [arXiv e-prints] {10.48550/arXiv.2206.05595}, \href {https://ui.adsabs.harvard.edu/abs/2022arXiv220605595G} {p. arXiv:2206.05595}

\bibitem[\protect\citeauthoryear{{Geller}, {Latham}  \& {Mathieu}}{{Geller} et~al.}{2015}]{2015AJ....150...97G}
{Geller} A.~M.,  {Latham} D.~W.,   {Mathieu} R.~D.,  2015, \mn@doi [\aj] {10.1088/0004-6256/150/3/97}, \href {https://ui.adsabs.harvard.edu/abs/2015AJ....150...97G} {150, 97}

\bibitem[\protect\citeauthoryear{{Geller}, {Mathieu}, {Latham}, {Pollack}, {Torres}  \& {Leiner}}{{Geller} et~al.}{2021}]{2021AJ....161..190G}
{Geller} A.~M.,  {Mathieu} R.~D.,  {Latham} D.~W.,  {Pollack} M.,  {Torres} G.,   {Leiner} E.~M.,  2021, \mn@doi [\aj] {10.3847/1538-3881/abdd23}, \href {https://ui.adsabs.harvard.edu/abs/2021AJ....161..190G} {161, 190}

\bibitem[\protect\citeauthoryear{{Gilliland} et~al.,}{{Gilliland} et~al.}{1993}]{1993AJ....106.2441G}
{Gilliland} R.~L.,  et~al., 1993, \mn@doi [\aj] {10.1086/116814}, \href {https://ui.adsabs.harvard.edu/abs/1993AJ....106.2441G} {106, 2441}

\bibitem[\protect\citeauthoryear{{Green}}{{Green}}{2018}]{2018JOSS....3..695M}
{Green} G.,  2018, \mn@doi [The Journal of Open Source Software] {10.21105/joss.00695}, \href {https://ui.adsabs.harvard.edu/abs/2018JOSS....3..695M} {3, 695}

\bibitem[\protect\citeauthoryear{{Green}, {Schlafly}, {Zucker}, {Speagle}  \& {Finkbeiner}}{{Green} et~al.}{2019}]{2019ApJ...887...93G}
{Green} G.~M.,  {Schlafly} E.,  {Zucker} C.,  {Speagle} J.~S.,   {Finkbeiner} D.,  2019, \mn@doi [\apj] {10.3847/1538-4357/ab5362}, \href {https://ui.adsabs.harvard.edu/abs/2019ApJ...887...93G} {887, 93}

\bibitem[\protect\citeauthoryear{{Grevesse} \& {Sauval}}{{Grevesse} \& {Sauval}}{1998}]{1998SSRv...85..161G}
{Grevesse} N.,  {Sauval} A.~J.,  1998, \mn@doi [\ssr] {10.1023/A:1005161325181}, \href {https://ui.adsabs.harvard.edu/abs/1998SSRv...85..161G} {85, 161}

\bibitem[\protect\citeauthoryear{{Herwig}}{{Herwig}}{2000}]{2000A&A...360..952H}
{Herwig} F.,  2000, \mn@doi [\aap] {10.48550/arXiv.astro-ph/0007139}, \href {https://ui.adsabs.harvard.edu/abs/2000A&A...360..952H} {360, 952}

\bibitem[\protect\citeauthoryear{{Holtzman} et~al.,}{{Holtzman} et~al.}{2018}]{2018AJ....156..125H}
{Holtzman} J.~A.,  et~al., 2018, \mn@doi [\aj] {10.3847/1538-3881/aad4f9}, \href {https://ui.adsabs.harvard.edu/abs/2018AJ....156..125H} {156, 125}

\bibitem[\protect\citeauthoryear{{Huber}, {Bryson}  \& {et al.}}{{Huber} et~al.}{2017}]{2017yCat.4034....0H}
{Huber} D.,  {Bryson} S.~T.,   {et al.} 2017, VizieR Online Data Catalog, \href {https://ui.adsabs.harvard.edu/abs/2017yCat.4034....0H} {p. IV/34}

\bibitem[\protect\citeauthoryear{{Iglesias} \& {Rogers}}{{Iglesias} \& {Rogers}}{1993}]{1993ApJ...412..752I}
{Iglesias} C.~A.,  {Rogers} F.~J.,  1993, \mn@doi [\apj] {10.1086/172958}, \href {https://ui.adsabs.harvard.edu/abs/1993ApJ...412..752I} {412, 752}

\bibitem[\protect\citeauthoryear{{Iglesias} \& {Rogers}}{{Iglesias} \& {Rogers}}{1996}]{1996ApJ...464..943I}
{Iglesias} C.~A.,  {Rogers} F.~J.,  1996, \mn@doi [\apj] {10.1086/177381}, \href {https://ui.adsabs.harvard.edu/abs/1996ApJ...464..943I} {464, 943}

\bibitem[\protect\citeauthoryear{{Irwin}}{{Irwin}}{2004}]{FreeEOS}
{Irwin} A.~W.,  2004, The FreeEOS Code for Calculating the Equation of State for Stellar Interiors, \url {http://freeeos.sourceforge.net/}

\bibitem[\protect\citeauthoryear{{Jermyn}, {Schwab}, {Bauer}, {Timmes}  \& {Potekhin}}{{Jermyn} et~al.}{2021}]{2021ApJ...913...72J}
{Jermyn} A.~S.,  {Schwab} J.,  {Bauer} E.,  {Timmes} F.~X.,   {Potekhin} A.~Y.,  2021, \mn@doi [\apj] {10.3847/1538-4357/abf48e}, \href {https://ui.adsabs.harvard.edu/abs/2021ApJ...913...72J} {913, 72}

\bibitem[\protect\citeauthoryear{{Jermyn} et~al.,}{{Jermyn} et~al.}{2023}]{2023ApJS..265...15J}
{Jermyn} A.~S.,  et~al., 2023, \mn@doi [\apjs] {10.3847/1538-4365/acae8d}, \href {https://ui.adsabs.harvard.edu/abs/2023ApJS..265...15J} {265, 15}

\bibitem[\protect\citeauthoryear{{Joyce} \& {Tayar}}{{Joyce} \& {Tayar}}{2023}]{2023Galax..11...75J}
{Joyce} M.,  {Tayar} J.,  2023, \mn@doi [Galaxies] {10.3390/galaxies11030075}, \href {https://ui.adsabs.harvard.edu/abs/2023Galax..11...75J} {11, 75}

\bibitem[\protect\citeauthoryear{Karakas}{Karakas}{2017}]{Karakas2017}
Karakas A.~I.,  2017, Low- and Intermediate-Mass Stars.
Springer International Publishing, Cham, pp 1--21, \mn@doi{10.1007/978-3-319-20794-0_117-1}, \url {https://doi.org/10.1007/978-3-319-20794-0_117-1}

\bibitem[\protect\citeauthoryear{{Khan} et~al.,}{{Khan} et~al.}{2019}]{2019A&A...628A..35K}
{Khan} S.,  et~al., 2019, \mn@doi [\aap] {10.1051/0004-6361/201935304}, \href {https://ui.adsabs.harvard.edu/abs/2019A&A...628A..35K} {628, A35}

\bibitem[\protect\citeauthoryear{{Kippenhahn}, {Weigert}  \& {Weiss}}{{Kippenhahn} et~al.}{2013}]{2013sse..book.....K}
{Kippenhahn} R.,  {Weigert} A.,   {Weiss} A.,  2013, {Stellar Structure and Evolution}, \mn@doi{10.1007/978-3-642-30304-3.
}

\bibitem[\protect\citeauthoryear{{Lebreton}, {Fernandes}  \& {Lejeune}}{{Lebreton} et~al.}{2001}]{2001A&A...374..540L}
{Lebreton} Y.,  {Fernandes} J.,   {Lejeune} T.,  2001, \mn@doi [\aap] {10.1051/0004-6361:20010757}, \href {https://ui.adsabs.harvard.edu/abs/2001A&A...374..540L} {374, 540}

\bibitem[\protect\citeauthoryear{{Leiner}, {Mathieu}, {Vanderburg}, {Gosnell}  \& {Smith}}{{Leiner} et~al.}{2019}]{2019ApJ...881...47L}
{Leiner} E.,  {Mathieu} R.~D.,  {Vanderburg} A.,  {Gosnell} N.~M.,   {Smith} J.~C.,  2019, \mn@doi [\apj] {10.3847/1538-4357/ab2bf8}, \href {https://ui.adsabs.harvard.edu/abs/2019ApJ...881...47L} {881, 47}

\bibitem[\protect\citeauthoryear{{Li} et~al.,}{{Li} et~al.}{2023}]{2023arXiv231116991L}
{Li} G.,  et~al., 2023, \mn@doi [arXiv e-prints] {10.48550/arXiv.2311.16991}, \href {https://ui.adsabs.harvard.edu/abs/2023arXiv231116991L} {p. arXiv:2311.16991}

\bibitem[\protect\citeauthoryear{{Li}, {Bi}, {Davies}, {Bedding}, {Li}, {Stello}  \& {Reyes}}{{Li} et~al.}{2024}]{2024MNRAS.530.2810L}
{Li} T.,  {Bi} S.,  {Davies} G.~R.,  {Bedding} T.~R.,  {Li} Y.,  {Stello} D.,   {Reyes} C.,  2024, \mn@doi [\mnras] {10.1093/mnras/stae1026}, \href {https://ui.adsabs.harvard.edu/abs/2024MNRAS.530.2810L} {530, 2810}

\bibitem[\protect\citeauthoryear{{Lindegren} et~al.,}{{Lindegren} et~al.}{2018}]{2018A&A...616A...2L}
{Lindegren} L.,  et~al., 2018, \mn@doi [\aap] {10.1051/0004-6361/201832727}, \href {https://ui.adsabs.harvard.edu/abs/2018A&A...616A...2L} {616, A2}

\bibitem[\protect\citeauthoryear{{Lindegren} et~al.,}{{Lindegren} et~al.}{2021}]{2021A&A...649A...4L}
{Lindegren} L.,  et~al., 2021, \mn@doi [\aap] {10.1051/0004-6361/202039653}, \href {https://ui.adsabs.harvard.edu/abs/2021A&A...649A...4L} {649, A4}

\bibitem[\protect\citeauthoryear{{Lindsay}, {Ong}  \& {Basu}}{{Lindsay} et~al.}{2024}]{2024arXiv240212461L}
{Lindsay} C.~J.,  {Ong} J.~M.~J.,   {Basu} S.,  2024, \mn@doi [arXiv e-prints] {10.48550/arXiv.2402.12461}, \href {https://ui.adsabs.harvard.edu/abs/2024arXiv240212461L} {p. arXiv:2402.12461}

\bibitem[\protect\citeauthoryear{{Liu}, {Asplund}, {Yong}, {Feltzing}, {Dotter}, {Mel{\'e}ndez}  \& {Ram{\'\i}rez}}{{Liu} et~al.}{2019}]{2019A&A...627A.117L}
{Liu} F.,  {Asplund} M.,  {Yong} D.,  {Feltzing} S.,  {Dotter} A.,  {Mel{\'e}ndez} J.,   {Ram{\'\i}rez} I.,  2019, \mn@doi [\aap] {10.1051/0004-6361/201935306}, \href {https://ui.adsabs.harvard.edu/abs/2019A&A...627A.117L} {627, A117}

\bibitem[\protect\citeauthoryear{{Magic}, {Serenelli}, {Weiss}  \& {Chaboyer}}{{Magic} et~al.}{2010}]{2010ApJ...718.1378M}
{Magic} Z.,  {Serenelli} A.,  {Weiss} A.,   {Chaboyer} B.,  2010, \mn@doi [\apj] {10.1088/0004-637X/718/2/1378}, \href {https://ui.adsabs.harvard.edu/abs/2010ApJ...718.1378M} {718, 1378}

\bibitem[\protect\citeauthoryear{{Marigo} \& {Aringer}}{{Marigo} \& {Aringer}}{2009}]{2009A&A...508.1539M}
{Marigo} P.,  {Aringer} B.,  2009, \mn@doi [\aap] {10.1051/0004-6361/200912598}, \href {https://ui.adsabs.harvard.edu/abs/2009A&A...508.1539M} {508, 1539}

\bibitem[\protect\citeauthoryear{{Mathieu} \& {Latham}}{{Mathieu} \& {Latham}}{1986}]{1986AJ.....92.1364M}
{Mathieu} R.~D.,  {Latham} D.~W.,  1986, \mn@doi [\aj] {10.1086/114269}, \href {https://ui.adsabs.harvard.edu/abs/1986AJ.....92.1364M} {92, 1364}

\bibitem[\protect\citeauthoryear{{Mathieu}, {van den Berg}, {Torres}, {Latham}, {Verbunt}  \& {Stassun}}{{Mathieu} et~al.}{2003}]{2003AJ....125..246M}
{Mathieu} R.~D.,  {van den Berg} M.,  {Torres} G.,  {Latham} D.,  {Verbunt} F.,   {Stassun} K.,  2003, \mn@doi [\aj] {10.1086/344944}, \href {https://ui.adsabs.harvard.edu/abs/2003AJ....125..246M} {125, 246}

\bibitem[\protect\citeauthoryear{{Mathys}}{{Mathys}}{1991}]{1991A&A...245..467M}
{Mathys} G.,  1991, \aap, \href {https://ui.adsabs.harvard.edu/abs/1991A&A...245..467M} {245, 467}

\bibitem[\protect\citeauthoryear{{Miglio} et~al.,}{{Miglio} et~al.}{2012}]{2012MNRAS.419.2077M}
{Miglio} A.,  et~al., 2012, \mn@doi [\mnras] {10.1111/j.1365-2966.2011.19859.x}, \href {https://ui.adsabs.harvard.edu/abs/2012MNRAS.419.2077M} {419, 2077}

\bibitem[\protect\citeauthoryear{{Montgomery}, {Marschall}  \& {Janes}}{{Montgomery} et~al.}{1993}]{1993AJ....106..181M}
{Montgomery} K.~A.,  {Marschall} L.~A.,   {Janes} K.~A.,  1993, \mn@doi [\aj] {10.1086/116628}, \href {https://ui.adsabs.harvard.edu/abs/1993AJ....106..181M} {106, 181}

\bibitem[\protect\citeauthoryear{{Morton}}{{Morton}}{2015}]{2015ascl.soft03010M}
{Morton} T.~D.,  2015, {isochrones: Stellar model grid package}, Astrophysics Source Code Library, record ascl:1503.010 (\mn@eprint {ascl} {1503.010})

\bibitem[\protect\citeauthoryear{{Mosumgaard}, {Ball}, {Silva Aguirre}, {Weiss}  \& {Christensen-Dalsgaard}}{{Mosumgaard} et~al.}{2018}]{2018MNRAS.478.5650M}
{Mosumgaard} J.~R.,  {Ball} W.~H.,  {Silva Aguirre} V.,  {Weiss} A.,   {Christensen-Dalsgaard} J.,  2018, \mn@doi [\mnras] {10.1093/mnras/sty1442}, \href {https://ui.adsabs.harvard.edu/abs/2018MNRAS.478.5650M} {478, 5650}

\bibitem[\protect\citeauthoryear{{Nguyen} et~al.,}{{Nguyen} et~al.}{2022}]{2022A&A...665A.126N}
{Nguyen} C.~T.,  et~al., 2022, \mn@doi [\aap] {10.1051/0004-6361/202244166}, \href {https://ui.adsabs.harvard.edu/abs/2022A&A...665A.126N} {665, A126}

\bibitem[\protect\citeauthoryear{{Nissen}, {Twarog}  \& {Crawford}}{{Nissen} et~al.}{1987}]{1987AJ.....93..634N}
{Nissen} P.~E.,  {Twarog} B.~A.,   {Crawford} D.~L.,  1987, \mn@doi [\aj] {10.1086/114345}, \href {https://ui.adsabs.harvard.edu/abs/1987AJ.....93..634N} {93, 634}

\bibitem[\protect\citeauthoryear{{Paxton}, {Bildsten}, {Dotter}, {Herwig}, {Lesaffre}  \& {Timmes}}{{Paxton} et~al.}{2011}]{2011ApJS..192....3P}
{Paxton} B.,  {Bildsten} L.,  {Dotter} A.,  {Herwig} F.,  {Lesaffre} P.,   {Timmes} F.,  2011, \mn@doi [\apjs] {10.1088/0067-0049/192/1/3}, \href {https://ui.adsabs.harvard.edu/abs/2011ApJS..192....3P} {192, 3}

\bibitem[\protect\citeauthoryear{{Paxton} et~al.,}{{Paxton} et~al.}{2013}]{2013ApJS..208....4P}
{Paxton} B.,  et~al., 2013, \mn@doi [\apjs] {10.1088/0067-0049/208/1/4}, \href {https://ui.adsabs.harvard.edu/abs/2013ApJS..208....4P} {208, 4}

\bibitem[\protect\citeauthoryear{{Paxton} et~al.,}{{Paxton} et~al.}{2015}]{2015ApJS..220...15P}
{Paxton} B.,  et~al., 2015, \mn@doi [\apjs] {10.1088/0067-0049/220/1/15}, \href {https://ui.adsabs.harvard.edu/abs/2015ApJS..220...15P} {220, 15}

\bibitem[\protect\citeauthoryear{{Paxton} et~al.,}{{Paxton} et~al.}{2018}]{2018ApJS..234...34P}
{Paxton} B.,  et~al., 2018, \mn@doi [\apjs] {10.3847/1538-4365/aaa5a8}, \href {https://ui.adsabs.harvard.edu/abs/2018ApJS..234...34P} {234, 34}

\bibitem[\protect\citeauthoryear{{Paxton} et~al.,}{{Paxton} et~al.}{2019}]{2019ApJS..243...10P}
{Paxton} B.,  et~al., 2019, \mn@doi [\apjs] {10.3847/1538-4365/ab2241}, \href {https://ui.adsabs.harvard.edu/abs/2019ApJS..243...10P} {243, 10}

\bibitem[\protect\citeauthoryear{{Pinsonneault}, {Terndrup}, {Hanson}  \& {Stauffer}}{{Pinsonneault} et~al.}{2004}]{2004ApJ...600..946P}
{Pinsonneault} M.~H.,  {Terndrup} D.~M.,  {Hanson} R.~B.,   {Stauffer} J.~R.,  2004, \mn@doi [\apj] {10.1086/379925}, \href {https://ui.adsabs.harvard.edu/abs/2004ApJ...600..946P} {600, 946}

\bibitem[\protect\citeauthoryear{{Planck Collaboration} et~al.,}{{Planck Collaboration} et~al.}{2016}]{2016A&A...594A..13P}
{Planck Collaboration} et~al., 2016, \mn@doi [\aap] {10.1051/0004-6361/201525830}, \href {https://ui.adsabs.harvard.edu/abs/2016A&A...594A..13P} {594, A13}

\bibitem[\protect\citeauthoryear{{Reimers}}{{Reimers}}{1975}]{1975psae.book..229R}
{Reimers} D.,  1975, in , Problems in stellar atmospheres and envelopes..
Springer-Verlag, pp 229--256

\bibitem[\protect\citeauthoryear{{Rogers} \& {Nayfonov}}{{Rogers} \& {Nayfonov}}{2002}]{2002ApJ...576.1064R}
{Rogers} F.~J.,  {Nayfonov} A.,  2002, \mn@doi [\apj] {10.1086/341894}, \href {https://ui.adsabs.harvard.edu/abs/2002ApJ...576.1064R} {576, 1064}

\bibitem[\protect\citeauthoryear{{Roxburgh}}{{Roxburgh}}{1992}]{1992A&A...266..291R}
{Roxburgh} I.~W.,  1992, \aap, \href {https://ui.adsabs.harvard.edu/abs/1992A&A...266..291R} {266, 291}

\bibitem[\protect\citeauthoryear{{Salaris} \& {Cassisi}}{{Salaris} \& {Cassisi}}{2017}]{2017RSOS....470192S}
{Salaris} M.,  {Cassisi} S.,  2017, \mn@doi [Royal Society Open Science] {10.1098/rsos.170192}, \href {https://ui.adsabs.harvard.edu/abs/2017RSOS....470192S} {4, 170192}

\bibitem[\protect\citeauthoryear{{Sanders}}{{Sanders}}{1977}]{1977A&AS...27...89S}
{Sanders} W.~L.,  1977, \aaps, \href {https://ui.adsabs.harvard.edu/abs/1977A&AS...27...89S} {27, 89}

\bibitem[\protect\citeauthoryear{{Sandquist}}{{Sandquist}}{2004}]{2004MNRAS.347..101S}
{Sandquist} E.~L.,  2004, \mn@doi [\mnras] {10.1111/j.1365-2966.2004.07174.x}, \href {https://ui.adsabs.harvard.edu/abs/2004MNRAS.347..101S} {347, 101}

\bibitem[\protect\citeauthoryear{{Sandquist} et~al.,}{{Sandquist} et~al.}{2016}]{2016ApJ...831...11S}
{Sandquist} E.~L.,  et~al., 2016, \mn@doi [\apj] {10.3847/0004-637X/831/1/11}, \href {https://ui.adsabs.harvard.edu/abs/2016ApJ...831...11S} {831, 11}

\bibitem[\protect\citeauthoryear{{Sandquist} et~al.,}{{Sandquist} et~al.}{2021}]{2021AJ....161...59S}
{Sandquist} E.~L.,  et~al., 2021, \mn@doi [\aj] {10.3847/1538-3881/abca8d}, \href {https://ui.adsabs.harvard.edu/abs/2021AJ....161...59S} {161, 59}

\bibitem[\protect\citeauthoryear{{Sarajedini}, {Dotter}  \& {Kirkpatrick}}{{Sarajedini} et~al.}{2009}]{2009ApJ...698.1872S}
{Sarajedini} A.,  {Dotter} A.,   {Kirkpatrick} A.,  2009, \mn@doi [\apj] {10.1088/0004-637X/698/2/1872}, \href {https://ui.adsabs.harvard.edu/abs/2009ApJ...698.1872S} {698, 1872}

\bibitem[\protect\citeauthoryear{{Saumon}, {Chabrier}  \& {van Horn}}{{Saumon} et~al.}{1995}]{1995ApJS...99..713S}
{Saumon} D.,  {Chabrier} G.,   {van Horn} H.~M.,  1995, \mn@doi [\apjs] {10.1086/192204}, \href {https://ui.adsabs.harvard.edu/abs/1995ApJS...99..713S} {99, 713}

\bibitem[\protect\citeauthoryear{Schlafly \& Finkbeiner}{Schlafly \& Finkbeiner}{2011}]{Schlafly_2011}
Schlafly E.~F.,  Finkbeiner D.~P.,  2011, \mn@doi [The Astrophysical Journal] {10.1088/0004-637X/737/2/103}, 737, 103

\bibitem[\protect\citeauthoryear{{Schlegel}, {Finkbeiner}  \& {Davis}}{{Schlegel} et~al.}{1998}]{1998ApJ...500..525S}
{Schlegel} D.~J.,  {Finkbeiner} D.~P.,   {Davis} M.,  1998, \mn@doi [\apj] {10.1086/305772}, \href {https://ui.adsabs.harvard.edu/abs/1998ApJ...500..525S} {500, 525}

\bibitem[\protect\citeauthoryear{{Sch{\"o}nrich}, {McMillan}  \& {Eyer}}{{Sch{\"o}nrich} et~al.}{2019}]{2019MNRAS.487.3568S}
{Sch{\"o}nrich} R.,  {McMillan} P.,   {Eyer} L.,  2019, \mn@doi [\mnras] {10.1093/mnras/stz1451}, \href {https://ui.adsabs.harvard.edu/abs/2019MNRAS.487.3568S} {487, 3568}

\bibitem[\protect\citeauthoryear{{Song}, {Alexeeva}, {Sitnova}, {Wang}, {Grupp}  \& {Zhao}}{{Song} et~al.}{2020}]{2020A&A...635A.176S}
{Song} N.,  {Alexeeva} S.,  {Sitnova} T.,  {Wang} L.,  {Grupp} F.,   {Zhao} G.,  2020, \mn@doi [\aap] {10.1051/0004-6361/201937110}, \href {https://ui.adsabs.harvard.edu/abs/2020A&A...635A.176S} {635, A176}

\bibitem[\protect\citeauthoryear{{Souto} et~al.,}{{Souto} et~al.}{2018}]{2018ApJ...857...14S}
{Souto} D.,  et~al., 2018, \mn@doi [\apj] {10.3847/1538-4357/aab612}, \href {https://ui.adsabs.harvard.edu/abs/2018ApJ...857...14S} {857, 14}

\bibitem[\protect\citeauthoryear{{Souto} et~al.,}{{Souto} et~al.}{2019}]{2019ApJ...874...97S}
{Souto} D.,  et~al., 2019, \mn@doi [\apj] {10.3847/1538-4357/ab0b43}, \href {https://ui.adsabs.harvard.edu/abs/2019ApJ...874...97S} {874, 97}

\bibitem[\protect\citeauthoryear{{Spoo} et~al.,}{{Spoo} et~al.}{2022}]{2022AJ....163..229S}
{Spoo} T.,  et~al., 2022, \mn@doi [\aj] {10.3847/1538-3881/ac5d53}, \href {https://ui.adsabs.harvard.edu/abs/2022AJ....163..229S} {163, 229}

\bibitem[\protect\citeauthoryear{{Stassun} \& {Torres}}{{Stassun} \& {Torres}}{2018}]{2018ApJ...862...61S}
{Stassun} K.~G.,  {Torres} G.,  2018, \mn@doi [\apj] {10.3847/1538-4357/aacafc}, \href {https://ui.adsabs.harvard.edu/abs/2018ApJ...862...61S} {862, 61}

\bibitem[\protect\citeauthoryear{{Stassun} et~al.,}{{Stassun} et~al.}{2019}]{2019AJ....158..138S}
{Stassun} K.~G.,  et~al., 2019, \mn@doi [\aj] {10.3847/1538-3881/ab3467}, \href {https://ui.adsabs.harvard.edu/abs/2019AJ....158..138S} {158, 138}

\bibitem[\protect\citeauthoryear{{Stello} et~al.,}{{Stello} et~al.}{2006}]{2006MNRAS.373.1141S}
{Stello} D.,  et~al., 2006, \mn@doi [\mnras] {10.1111/j.1365-2966.2006.11060.x}, \href {https://ui.adsabs.harvard.edu/abs/2006MNRAS.373.1141S} {373, 1141}

\bibitem[\protect\citeauthoryear{{Stello} et~al.,}{{Stello} et~al.}{2016}]{2016ApJ...832..133S}
{Stello} D.,  et~al., 2016, \mn@doi [\apj] {10.3847/0004-637X/832/2/133}, \href {https://ui.adsabs.harvard.edu/abs/2016ApJ...832..133S} {832, 133}

\bibitem[\protect\citeauthoryear{{Strom}, {Strom}  \& {Bregman}}{{Strom} et~al.}{1971}]{1971PASP...83..768S}
{Strom} S.~E.,  {Strom} K.~M.,   {Bregman} J.~N.,  1971, \mn@doi [\pasp] {10.1086/129213}, \href {https://ui.adsabs.harvard.edu/abs/1971PASP...83..768S} {83, 768}

\bibitem[\protect\citeauthoryear{{Taylor}}{{Taylor}}{2007}]{2007AJ....133..370T}
{Taylor} B.~J.,  2007, \mn@doi [\aj] {10.1086/509781}, \href {https://ui.adsabs.harvard.edu/abs/2007AJ....133..370T} {133, 370}

\bibitem[\protect\citeauthoryear{{Timmes} \& {Swesty}}{{Timmes} \& {Swesty}}{2000}]{2000ApJS..126..501T}
{Timmes} F.~X.,  {Swesty} F.~D.,  2000, \mn@doi [\apjs] {10.1086/313304}, \href {https://ui.adsabs.harvard.edu/abs/2000ApJS..126..501T} {126, 501}

\bibitem[\protect\citeauthoryear{{Trampedach}, {Stein}, {Christensen-Dalsgaard}, {Nordlund}  \& {Asplund}}{{Trampedach} et~al.}{2014a}]{2014MNRAS.442..805T}
{Trampedach} R.,  {Stein} R.~F.,  {Christensen-Dalsgaard} J.,  {Nordlund} {\r{A}}.,   {Asplund} M.,  2014a, \mn@doi [\mnras] {10.1093/mnras/stu889}, \href {https://ui.adsabs.harvard.edu/abs/2014MNRAS.442..805T} {442, 805}

\bibitem[\protect\citeauthoryear{{Trampedach}, {Stein}, {Christensen-Dalsgaard}, {Nordlund}  \& {Asplund}}{{Trampedach} et~al.}{2014b}]{2014MNRAS.445.4366T}
{Trampedach} R.,  {Stein} R.~F.,  {Christensen-Dalsgaard} J.,  {Nordlund} {\r{A}}.,   {Asplund} M.,  2014b, \mn@doi [\mnras] {10.1093/mnras/stu2084}, \href {https://ui.adsabs.harvard.edu/abs/2014MNRAS.445.4366T} {445, 4366}

\bibitem[\protect\citeauthoryear{{VandenBerg} \& {Stetson}}{{VandenBerg} \& {Stetson}}{2004}]{2004PASP..116..997V}
{VandenBerg} D.~A.,  {Stetson} P.~B.,  2004, \mn@doi [\pasp] {10.1086/426340}, \href {https://ui.adsabs.harvard.edu/abs/2004PASP..116..997V} {116, 997}

\bibitem[\protect\citeauthoryear{{VandenBerg}, {Bergbusch}  \& {Dowler}}{{VandenBerg} et~al.}{2006}]{2006ApJS..162..375V}
{VandenBerg} D.~A.,  {Bergbusch} P.~A.,   {Dowler} P.~D.,  2006, \mn@doi [\apjs] {10.1086/498451}, \href {https://ui.adsabs.harvard.edu/abs/2006ApJS..162..375V} {162, 375}

\bibitem[\protect\citeauthoryear{{VandenBerg}, {Gustafsson}, {Edvardsson}, {Eriksson}  \& {Ferguson}}{{VandenBerg} et~al.}{2007}]{2007ApJ...666L.105V}
{VandenBerg} D.~A.,  {Gustafsson} B.,  {Edvardsson} B.,  {Eriksson} K.,   {Ferguson} J.,  2007, \mn@doi [\apjl] {10.1086/521877}, \href {https://ui.adsabs.harvard.edu/abs/2007ApJ...666L.105V} {666, L105}

\bibitem[\protect\citeauthoryear{{Viani} \& {Basu}}{{Viani} \& {Basu}}{2020}]{2020ApJ...904...22V}
{Viani} L.~S.,  {Basu} S.,  2020, \mn@doi [\apj] {10.3847/1538-4357/abba17}, \href {https://ui.adsabs.harvard.edu/abs/2020ApJ...904...22V} {904, 22}

\bibitem[\protect\citeauthoryear{{Yadav} et~al.,}{{Yadav} et~al.}{2008}]{2008A&A...484..609Y}
{Yadav} R.~K.~S.,  et~al., 2008, \mn@doi [\aap] {10.1051/0004-6361:20079245}, \href {https://ui.adsabs.harvard.edu/abs/2008A&A...484..609Y} {484, 609}

\bibitem[\protect\citeauthoryear{{Zinn}}{{Zinn}}{2021}]{2021AJ....161..214Z}
{Zinn} J.~C.,  2021, \mn@doi [\aj] {10.3847/1538-3881/abe936}, \href {https://ui.adsabs.harvard.edu/abs/2021AJ....161..214Z} {161, 214}

\bibitem[\protect\citeauthoryear{{Zinn}, {Pinsonneault}, {Huber}  \& {Stello}}{{Zinn} et~al.}{2019}]{2019ApJ...878..136Z}
{Zinn} J.~C.,  {Pinsonneault} M.~H.,  {Huber} D.,   {Stello} D.,  2019, \mn@doi [\apj] {10.3847/1538-4357/ab1f66}, \href {https://ui.adsabs.harvard.edu/abs/2019ApJ...878..136Z} {878, 136}

\makeatother
\end{thebibliography}




\appendix


\bsp	
\label{lastpage}
\end{document}